\documentclass{aa}

\usepackage{graphicx}
\usepackage{url}

\newcommand{\EQ}{\begin{equation}}
\newcommand{\EN}{\end{equation}}

\newcommand{\Eq}[1]{Eq.~(\ref{#1})}
\newcommand{\uu}{{\vec{u}}}
\newcommand{\BB}{{\vec{B}}}
\newcommand{\JJ}{{\vec{J}}}
\newcommand{\EE}{{\vec{E}}}
\newcommand{\FF}{{\vec{F}}}
\newcommand{\AAA}{{\vec{A}}}
\newcommand{\SSS}{{\vec{S}}}
\newcommand{\ttau}{{\vec{\tau}}}
\newcommand{\half}{{\textstyle{1\over2}}}

\newcommand{\dd}{{\rm d} {}}
\newcommand{\DD}{{\rm D} {}}

\newcommand{\km}{\,{\rm km}}
\newcommand{\s}{\,{\rm s}}
\newcommand{\K}{\,{\rm K}}
\newcommand{\G}{\,{\rm G}}
\newcommand{\AU}{\,{\rm AU}}
\newcommand{\g}{\,{\rm g}}
\newcommand{\cm}{\,{\rm cm}}
\newcommand{\yr}{\,{\rm yr}}

\newcommand{\cs}{{c_{\rm s}}}

\usepackage{times}

\begin{document}


\title{Structured outflow from a dynamo active accretion disc}

\author{
  B.~von~Rekowski\thanks{Now at: Department of Astronomy \& Space Physics,
  Uppsala University, Box 515, 751 20 Uppsala, Sweden}
  \and A.~Brandenburg\thanks{NORDITA, Blegdamsvej 17,
  DK-2100 Copenhagen \O, Denmark}
  \and W.~Dobler\thanks{Kiepenheuer Institute for Solar Physics,
  Sch\"oneckstr. 6, 79104 Freiburg, Germany}
  \and A.~Shukurov
}

\institute{
  Department of Mathematics, University of Newcastle,
  Newcastle upon Tyne NE1~7RU, UK
}

\offprints{Brigitta.vonRekowski@astro.uu.se}

\date{\today,~ $ $Revision: 1.105.2.28 $ $}



\abstract{
  We present an axisymmetric numerical model of a dynamo active accretion disc.
  If the dynamo-generated magnetic field in the disc is sufficiently strong
   (close to equipartition with thermal energy),
 a fast magneto-centrifugally driven outflow develops within a conical
 shell near the
  rotation axis, together with
  a slower pressure driven outflow from the outer
  parts of the disc as well as around the axis. Our results show that a dynamo
  active accretion disc can
  contribute to driving an outflow even without any external magnetic field.
 The fast outflow in the conical shell is confined by the
  azimuthal field
  which is produced by the dynamo in the disc and advected to the disc
  corona. This part of the outflow has high angular momentum and
  is cooler and less dense than its surroundings.
  The conical shell's half-opening angle is typically about $30^\circ$
  near the disc and decreases slightly with height.
  The slow outflow is hotter and denser.
}
\maketitle

\section{Introduction}
The importance of magnetic fields for disc accretion is widely recognized
(e.g.,\ Mestel 1999), and the turbulent dynamo is believed to be a major
source of magnetic fields in accretion discs (Pudritz 1981a, 1981b; Stepinski
\& Levy 1988; Brandenburg et al.\ 1995; Hawley et al.\ 1996).
A magnetic field anchored in the disc is further considered to be a
major factor in launching and  collimating a wind in young
stellar objects and active galactic nuclei (Blandford \& Payne 1982;
Pelletier \& Pudritz 1992); see K\"onigl \& Pudritz
(2000) for a recent review of stellar outflows. Yet, most models of the
formation and collimation of jets rely on an externally imposed
poloidal magnetic field and disregard any field produced in the disc.
Our aim here is to study outflows in connection with dynamo-generated
magnetic fields.
We discuss parameters of young stellar objects in our estimates, but the
model also applies to systems containing a compact central object after rescaling
and possibly other modifications such as the appropriate choice
of the gravitational potential.

Extensive numerical studies of collimated disc winds have been performed
using several types of model. Uchida \& Shibata (1985), Matsumoto et al.\
(1996) and Kudoh et al.\ (1998) consider an ideal MHD model
of an accretion disc embedded in a non-rotating corona, permeated by
an external magnetic field initially aligned with the disc rotation axis.
Intense accretion develops in the disc (Stone \& Norman 1994), accompanied
by a centrifugally driven outflow. The wind is eventually collimated by
toroidal magnetic field produced in the corona by winding up the poloidal
field.

Bell (1994), Bell \& Lucek (1995) and Lucek \& Bell (1996, 1997)
use two- and three-dimensional ideal MHD models, with a polytropic equation
of state, to study the formation and stability of pressure driven jets
collimated by poloidal magnetic field, which has a minimum inside the jet.
The general structure of thermally driven disc winds was studied analytically
by, e.g.,\ Fukue (1989) and Takahara et al.\ (1989).

Another type of model was developed by Ustyugova et al.\ (1995), Romanova
et al.\ (1997, 1998), Ouyed et al.\ (1997) and Ouyed \& Pudritz
(1997a, 1997b, 1999) who consider ideal MHD in a polytropic corona permeated by
an external poloidal magnetic field and subsume the physics of the
accretion disc into boundary conditions at the base of the corona.
The disc is assumed to be in Keplerian rotation (with any accretion
neglected). The corona is
non-rotating initially, and the system is driven by the injection of matter
through the boundary that represents the disc surface.
These models develop a steady (or at least statistically steady)
state consistent with the analytical models by Blandford \& Payne (1982),
showing a magneto-centrifugal wind collimated by toroidal magnetic field,
which again is produced in the corona by the vertical shear in the angular
velocity.
Three-dimensional simulations suggest that the resulting collimated
outflow does not break due to the kink instability (Ouyed et al.\ 2003;
Thomsen \& Nordlund 2003).

It is not quite clear how strong the external magnetic field in accretion
discs can be. Dragging
of an external field from large radii in a viscous disc requires that magnetic diffusivity is
much smaller than the kinetic one (Lubow et al.\ 1994), i.e.\ that the magnetic Prandtl number
is significantly larger than unity
which would be difficult to explain in a turbulent disc (Heyvaerts et al.\ 1996).
This argument neglects, however, the effect of magnetic torques which could
produce significant field line dragging even when the magnetic Prandtl number
is of order unity (Shalybkov \& R\"udiger 2000). On the other hand,
the efficiency of trapping
an external magnetic field at initial stages of the disc formation is questionable because
only a small fraction of the external flux can be retained if the density contrast between
the disc and the surrounding medium is large. In that case the magnetic field will be strongly bent
and reconnection will remove most of the flux from the disc. Furthermore,
the ionization fraction is probably too small
in the inner disc to ensure sufficient
coupling between the gas and the magnetic field (Fromang et al.\ 2002).
For these reasons it seems
appropriate to explore whether or not magnetic fields generated by a dynamo within
the disc can produce an outflow with realistic properties.

The models of Ustyugova et al.\ (1995) and Ouyed \& Pudritz (1997a, 1997b, 1999) have a
distributed mass source and imposed poloidal velocity at the (unresolved) disc surface. As a result,
their system develops a persistent outflow which is not just a transient. On the other hand, the models of
Matsumoto et al.\ (1996) resolve the disc, but have no mass replenishment to compensate losses
via the outflow and accretion, and so the disc disappears with time. Our model is an attempt to
combine advantages of both of these approaches and also to add dynamo action in the disc.
Instead of a rigidly prescribed mass injection, we allow for
self-regulatory replenishment of matter within a resolved disc. Instead of prescribing
poloidal velocity at the disc surface, we resolve the disc and prescribe an
entropy contrast between the disc and
the corona, leaving more freedom for the system. Such an increase of entropy with
height is only natural to expect for a disc surrounded by a hot corona,
and we
parameterize the coronal heating by a (fixed) entropy contrast. We further add self-sustained, intrinsic
magnetic field to our system, as opposed to an external field used in the other models.
Since our model goes beyond ideal MHD, the magnetic field must be maintained against
decay. A simple form of mean-field dynamo action is included for this
purpose.

Like in the model of Ouyed \& Pudritz (1997a, 1997b, 1999), the hot, pressure supported corona does
not rotate initially. The disc is cool and is therefore centrifugally supported, so its
rotation is nearly Keplerian. The corresponding steady-state solution is used as our
initial condition. This solution is, however, unstable because of the vertical shear in
the angular velocity between the disc and the corona (Urpin \& Brandenburg 1998).
Angular momentum transfer by viscous
and magnetic
stresses also leads to a departure from the initial state.
As a
result, a meridional flow develops, which exchanges matter between the corona and the
disc surface layers. Mass losses through the disc surface and to the accreting central object
are then replenished in
the disc where we allow local mass production whenever and wherever the
density decreases below a reference value.
Matter is heated as it moves into the corona; this leads to
an increase in pressure which drives the wind.
Another efficient driver of the outflow in our model is
magneto-centrifugal acceleration.
A strong toroidal magnetic field produced by the
dynamo in the disc is advected into the corona and contributes to confining the wind.

Altogether, our model contains many features included in a range of earlier models.  For example,
the disc is essentially resistive to admit dynamo action, whereas magnetic diffusion turns out to be
relatively unimportant in the corona, similarly to the
models of, e.g.,\ Wardle \& K\"onigl (1993), Ferreira \& Pelletier (1995) and Casse \& Ferreira
(2000a, 2000b).
The synthetic
nature of the outflow, driven by both pressure forces and magneto-centrifugally, is characteristic
of the self-similar solutions of  Ferreira \& Pelletier (1995) and Casse \& Ferreira (2000b).
The latter authors also stress the r\^ole of the vertical entropy gradient in enhancing the
mass flux in the outflow and prescribe a vertical profile of the entropy production rate.
Structured outflows outwardly similar to our results have been discussed by Krasnopolsky
et al.\ (1999) and Goodson et al.\ (1997).
However, in the former paper
the structure in their solutions is due to
the boundary conditions imposed on the disc surface, and in the latter model the structure is due to
the interaction between the stellar magnetic field
and the disc under the presence of differential rotation between star and disc.
On the contrary, the structure in our model
results from the dominance of different driving mechanisms in the inner and outer parts of the disc.

\begin{figure}
  \centering
  \includegraphics[height=5cm]{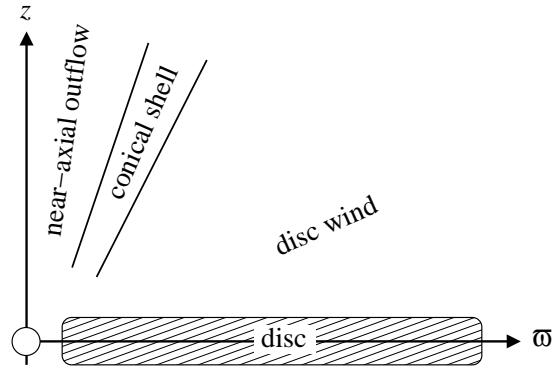}
  \caption{General structure of the outflows typically obtained in the
    magnetic runs with mass sink presented in this paper.
    The cool, dense disc emits
    (a) a hardly collimated, thermally driven wind
    (slow, hot, dense, magnetized, rotating),
    (b) a magneto-centrifugally driven outflow in a conical shell
    (faster, cooler, less dense, magnetized, quickly rotating),
    and
    (c) a thermally driven outflow near the axis
    (slow, hot, dense, weakly magnetized and weakly rotating).
  }
  \label{Fig-structure}
\end{figure}

While we had initially expected to find collimated outflows similar to
those reported by  Ouyed \& Pudritz (1997a, 1997b), the outflow patterns
we obtained turned out to be of quite different nature.
However, a significant difference in our model is the resolved disc,
which is necessary to model dynamo action in the disc, and the relatively small
extension into the corona, where collimation is not yet expected.
Figure~\ref{Fig-structure} shows a sketch of the overall, multi-component
structure of the outflows in the presence of magnetic fields and mass sink,
which needs
to be compared to the individual Figures in the rest of the paper.

\bigskip

The plan of the paper is as follows. We introduce our model in Sect.~\ref{TM}, and then consider
a range of parameters in Sect.~\ref{RS} to clarify and illustrate the main physical effects.
In Sect.~\ref{FinalModel} we explore the parameter space, and our conclusions are presented
in Sect.~\ref{Discu}.

\section{The model} \label{TM}

\subsection{Basic equations}

Our intention is to make our model as simple as possible and to avoid
detailed modelling of mass supply to the accretion disc, which occurs
differently in different accreting systems. For example,
matter enters the accretion disc
at large radii in a restricted range of azimuthal angles in binary systems
with Roche lobe overflow. On the other hand, matter supply
can be more uniform in both azimuth and radius in active galactic nuclei.
Being interested in other aspects of accretion physics, we prefer to avoid
detailed modelling of these processes. Instead,
similarly to the model of Ouyed \& Pudritz
(1997a, 1997b), we inject matter into the system, but an important difference is that
we introduce a self-regulating mass source in the disc.
We also allow for a mass sink at the centre to model
accretion onto the central object (star). With these two effects included,
the continuity equation becomes
\begin{equation}
{\partial\varrho\over\partial t}+\vec{\nabla}\cdot(\varrho\uu)
=q_\varrho^{\rm disc}+q_\varrho^{\rm star} \equiv \dot{\varrho},
\label{Cont}
\end{equation}
where $\uu$ is the velocity field and $\varrho$ is the gas density.
The mass source, $q_\varrho^{\rm
disc}$, is localized in the disc (excluding the star) and turns on once the local density
drops below a reference value $\varrho_0$ (chosen to be equal to the initial
density, see Sect.~\ref{Sec-initial}), i.e.\
\begin{equation}
q_\varrho^{\rm disc}=\frac{\xi_{\rm disc}(\vec{r})-\xi_{\rm star}(\vec{r})}
                          {\tau_{\rm disc}}\,(\varrho_0-\varrho)_+,
                                                \label{qrho_term}
\end{equation}
where the subscript `$+$' indicates that only positive values are used,
i.e.\
\begin{eqnarray}
x_+=
\left\{
\begin{array}{ll}
x\quad&\mbox{if $x>0$,}\\
0&\mbox{otherwise}.
\end{array}
\right.
\end{eqnarray}
This means that matter is injected only if
$\varrho<\varrho_0$, and that the strength of the mass source is
proportional to the gas density deficit.
Throughout this work we use cylindrical polar coordinates $(\varpi,\varphi,z)$,
assuming axisymmetry.
The shapes of the disc and the central object are specified with the profiles
$\xi_{\rm disc}$ and $\xi_{\rm star}$ via
\begin{eqnarray}
\xi_{\rm disc}(\varpi,z)&=&
\Theta\left({\varpi_0{-}\varpi},{d}\right)\,\Theta\left({z_0{-}|z|},{d}\right),
\label{xi}\\
\xi_{\rm star}(\varpi,z)&=&
\Theta\left({r_0{-}r},{d}\right).
\label{xistar}
\end{eqnarray}
Here, $\Theta(x,d)$ is a smoothed Heaviside step function
with a smoothing half-width $d$ set to 8 grid zones
in $\xi_{\rm disc}$ and $d=r_0$ in $\xi_{\rm star}$;
$r=\sqrt{\varpi^2+z^2}$ is the spherical radius and $r_0$ is the stellar
radius introduced in Eq.~(\ref{gravpot});
$\varpi_0$ and $z_0$ are the disc outer radius and disc half-thickness, respectively.
So, $\xi_{\rm disc}$ is equal to unity at the disc midplane and vanishes in
the corona, whereas $\xi_{\rm star}$ is unity at the centre of the central object
and vanishes outside the star.
Note that $\xi_{\rm disc}-\xi_{\rm star}\ge0$ everywhere, since
$\xi_{\rm disc}>0$) all the way to the origin,
$z=\varpi=0$.

In \Eq{qrho_term}, $\tau_{\rm disc}$ is a response time which is chosen to
be significantly shorter than the time scale of the depletion processes,
which is equivalent to
the time scale of mass replenishment in the disc, $M_{\rm disc}/\dot M_{\rm source}$
(cf.\ Sect.~\ref{M-E-loss}),
to avoid unphysical influences of the mass source. We do
not fix the distribution and magnitude of $q^{\rm disc}_\varrho$
beforehand, but the system adjusts itself such as to prevent the disc
from disappearing.

The self-regulating mass sink at the position of the central star
is modelled in a similar manner,
\begin{equation}
q_\varrho^{\rm star}
=-\frac{\xi_{\rm star}(\vec{r})}{\tau_{\rm star}}\,(\varrho-\varrho_0)_+,
                                                                        \label{qminus}
\end{equation}
where $\xi_{\rm star}$ is defined in Eq.~(\ref{xistar})
and $\tau_{\rm star}$
is a central accretion time scale that controls the efficiency of the sink.
We discuss physically meaningful values of $\tau_{\rm star}$ in Sect.~\ref{DVCP}.

Apart from the continuity equation (\ref{Cont}), the mass source also appears
in the Navier--Stokes equation, unless matter is always injected with the
ambient velocity of the gas. In that case, however, a runaway instability
can occur: if matter is already slower than Keplerian, it falls inward, and so
does the newly injected matter. This enhances the need for even more mass
injection. A similar argument applies also if matter is rotating faster
than Keplerian. This is why we inject matter at the Keplerian speed.
This leads to an extra term in the Navier--Stokes equation,
$(\vec{\uu}_{\rm K}-\vec{\uu}) q_\varrho^{\rm disc}$, which would only
be absent if the gas were rotating at the Keplerian speed. Thus,
the Navier--Stokes equation takes the form
\begin{equation}
    \frac{\DD\uu}{\DD t}
    =
    - {1\over\varrho}\vec{\nabla}p - \vec{\nabla}\Phi
    + \frac{1}{\varrho}\left[\vec{F} + (\vec{\uu}_{\rm K}-\vec{\uu}) q_\varrho^{\rm disc}\right],
\label{momentum}
\end{equation}
where ${\rm D}/{\rm D}t=\partial/\partial t+\uu\cdot\vec{\nabla}$ is
the advective derivative, $p$ is the gas pressure,
$\Phi$ is the gravitational potential,
$\FF=\JJ\times\BB+\vec{\nabla}\cdot\ttau$ is the sum of the Lorentz
and viscous forces, $\JJ=\vec{\nabla}\times\BB/\mu_0$ is the current density due to
the (mean) magnetic field $\BB$, $\mu_0$ is the magnetic permeability,
and $\ttau$
is the viscous stress tensor,
\EQ
\tau_{ij}=\varrho\nu(\partial u_i/\partial x_j+\partial u_j/\partial x_i).
\EN
Here, $\nu$ is the kinematic viscosity, which
has been subdivided into three contributions,
\EQ
\nu=\nu_{\rm t}+\nu_{\rm adv}+\nu_{\rm shock}.
\EN
The first term is a turbulent (Shakura--Sunyaev) viscosity in the disc,
\EQ
\nu_{\rm t}=\alpha_{\rm SS}c_{\rm s}z_0\xi_{\rm disc}(\vec{r}),
\label{nut}
\EN
where $c_{\rm s}=(\gamma p/\varrho)^{1/2}$ is the sound speed
and $\gamma$ is the ratio of specific heats.
The second term is an artificial advection viscosity,
\EQ
\nu_{\rm adv}=c_\nu^{\rm adv}\delta x\,(\uu_{\rm pol}^2+c_{\rm s}^2+v_{\rm A,pol}^2)^{1/2},
\label{nuadv}
\EN
which is required to stabilize rapidly moving patterns. Here,
$\delta x = \min(\delta\varpi,\delta z)$ is the mesh size,
and $c_\nu^{\rm adv}$ is a constant specified in Sect.~\ref{DVCP}.
In \Eq{nuadv} we have only used the poloidal velocity, $\uu_{\rm pol}$, and
the Alfv\'en speed due to the poloidal magnetic field,
$v_{\rm A,pol}=(\vec{B}_{\rm pol}^2/\varrho\mu_0)^{1/2}$, because in
an axisymmetric calculation advection in the $\varphi$-direction is unimportant.
Finally,
\EQ
\nu_{\rm shock}=c_\nu^{\rm shock}\delta x^2(-\vec{\nabla}\cdot\uu)_+
\EN
is an artificial shock viscosity,
with $c_\nu^{\rm shock}$ a constant specified in Sect.~\ref{DVCP}.
Note that
$\nu_{\rm adv}$ and $\nu_{\rm shock}$ are needed for numerical
reasons; they tend to zero for increasing
resolution, $\delta x \to 0$.


We assume that the magnetic field in the disc is generated by
a standard $\alpha^2\Omega$-dynamo (e.g.,\ Krause \& R\"adler 1980),
which implies an extra electromotive force, $\alpha\BB$, in the
induction equation for the mean magnetic field, $\BB$.
To ensure that $\BB$ is solenoidal, we solve the
induction equation in terms of the vector potential $\AAA$,
\begin{equation}
{\partial\AAA\over\partial t}=\uu\times\BB+\alpha\BB-\eta\mu_0\JJ,
\label{Ind}
\end{equation}
where $\BB=\vec{\nabla}\times\AAA$,
and $\eta$ is the magnetic diffusivity.

Since $\alpha(\vec{r})$ has to be
antisymmetric about the midplane and vanishing outside the disc,
we adopt the form
\begin{equation}
\alpha=\alpha_0\,{z\over
z_0}\,{\xi_{\rm disc}(\vec{r})\over1+v_{\rm A}^2/v_0^2},     \label{alpha}
\end{equation}
where $v_{\rm A}$ is the Alfv\'en speed based on the total magnetic field, and $\alpha_0$ and
$v_0$ are
parameters that control the intensity of dynamo action
and the field strength in the disc, respectively.
The $\alpha$-effect has been truncated near the axis, so that $\alpha=0$ for
$\varpi \le 0.2$.
For the magnetic diffusivity we assume
\begin{equation}
\eta=\eta_0+\eta_{\rm t},                     \label{eeta}
\end{equation}
where $\eta_0$ is a uniform background diffusivity
and the turbulent part, $\eta_{\rm t}=
\eta_{{\rm t}0}\xi_{\rm disc}(\vec{r})$, vanishes outside the disc.
Thus, magnetic diffusivity in the corona is smaller than in the disc.

Depending on the sign of $\alpha$ and the vertical distribution of $\eta$, the dynamo can
generate magnetic fields of either dipolar or quadrupolar symmetry.
We shall discuss both types of geometry.

\subsection{Implementation of a cool disc embedded in a hot corona}
\label{Sec-cool-hot}

Protostellar systems are known to be
strong X-ray sources (see, e.g.,\ Glassgold et al.\ 2000;
Feigelson \& Montmerle 1999; Grosso et al.\ 2000).
The X-ray emission is generally attributed to coronae of disc-star
systems, plausibly heated by small scale magnetic reconnection events
(Galeev et al.\ 1979), for example in the form of nanoflares
that are caused by slow footpoint motions (Parker 1994).
Heating of disc coronae by fluctuating magnetic
fields is indeed quite natural if one accepts that the disc
turbulence is caused by the magneto-rotational instability.
Estimates for the coronal temperatures of YSOs range from $10^6\K$
to $10^8\K$ (see, e.g.,\ Feigelson \& Montmerle 1999).
For the base of the disc corona, temperatures
of at least $8\times10^3$ are to be expected in order to explain
the observed mass loss rates (Kwan \& Tademaru 1995; Kwan 1997).
The discs, on the other hand, have typical temperatures of a few
$10^3\K$ (e.g.,\ Papaloizou \& Terquem 1999).

A simple way to implement a dense, relatively cool disc embedded in a rarefied, hot
corona without modelling
the detailed physics of coronal heating is to prescribe the
distribution of specific entropy, $s(\vec{r})$, such that $s$ is smaller
within the disc and larger in the corona.
For a perfect gas this implies $p=K\varrho^\gamma$ (in a dimensionless form), where
$K = e^{s/c_v}$ is a function of position
(here $p$ and $\varrho$ are gas pressure and
density, $\gamma=c_p/c_v$, and $c_p$ and $c_v$ are the
specific heats at constant pressure and constant volume, respectively).

We prescribe the polytrope parameter $K$ to be unity in the
corona and smaller in the disc, so we put
\begin{equation}
[K(\vec{r})]^{1/\gamma} = e^{s/c_p} = 1-(1{-}\beta)\xi_{\rm disc}(\vec{r}),
\label{Kdef}
\end{equation}
where $0<\beta<1$ is a free parameter with $\beta=e^{-\Delta s/c_p}$,
that controls the entropy contrast, $\Delta s>0$, between corona and disc.
We consider values of $\beta$ between 0.1 and 0.005, which yields
an entropy contrast, $\Delta s/c_p$, between 2.3 and 5.3.
The temperature ratio between disc and corona is roughly $\beta$;
see \Eq{hdisc}. Assuming pressure equilibrium between disc and corona,
and $p\propto\rho T$ for a perfect gas, the corresponding density ratio is $\beta^{-1}$.

\subsection{Formulation in terms of potential enthalpy}

In the present case it is advantageous to use the potential enthalpy,
\EQ
  H=h+\Phi,
\EN
as a variable. Here, $h$ is the specific enthalpy, $h = c_v T + p/\varrho = c_p T$
for a perfect gas (with constant specific heats), and $T$ is temperature.
Therefore, specific enthalpy $h$ is related
to $p$ and $\varrho$ via
$h = \gamma(\gamma{-}1)^{-1} p/\varrho$.
Specific entropy $s$ is related
to $p$ and $\varrho$ (up to an additive constant)
through $s=c_v\ln p-c_p\ln\varrho$ for a perfect gas.
We have therefore
\EQ
  \frac{\DD\ln h}{\DD t}
  = \frac{\DD\ln p}{\DD t} - \frac{\DD\ln\varrho}{\DD t}
  = (\gamma{-}1) \frac{\DD\ln\varrho}{\DD t}
    + \frac{\gamma}{c_p}\frac{\DD s}{\DD t} .
\EN
Since $s$ is independent of time,
$\DD s/\DD t=\uu\cdot\vec{\nabla} s$.
Together with \Eq{Cont}, this yields an evolution equation for $h$,
which can be written in terms of $H$ as
\EQ
\frac{\DD H}{\DD t}=\uu\cdot\vec{\nabla}\Phi+(\gamma-1)h
\left({\dot{\varrho}\over\varrho}-\vec{\nabla}\cdot\uu\right)
+ {\gamma h\over c_p}\uu\cdot\vec{\nabla}s.
\label{DHDt}
\EN
In the following we solve \Eq{DHDt} instead of \Eq{Cont}.
In terms of $h$, density and sound speed are given by
\begin{equation}
\varrho^{\gamma-1}=\frac{\gamma-1}{\gamma}\,\frac{h}{K},\quad
c_{\rm s}^2=(\gamma-1)h= \gamma\frac{p}{\varrho}.
\label{rrho}
\end{equation}
It proved necessary to include an artificial diffusion term in
Eq.~(\ref{DHDt}) with a diffusion coefficient proportional to $\nu$.

The first law of thermodynamics allows us to
express the pressure gradient in terms of
$h$ and $s$,
\EQ
-{1\over\varrho}\vec{\nabla}p=-\vec{\nabla}h+T\vec{\nabla} s,
\label{Tgrads_in_pgrad}
\EN
and with $T=h/c_p$ we obtain
\begin{equation}
    \frac{\DD\uu}{\DD t}
    =
    - \vec{\nabla}H + h\vec{\nabla}{s/c_p}
    + \frac{1}{\varrho}\left[\vec{F} + (\vec{\uu}_{\rm K}-\vec{\uu}) q_\varrho^{\rm disc}\right],
\label{momentumH}
\end{equation}
which now replaces \Eq{momentum}.

We use a softened, spherically symmetric gravitational potential of the form
\begin{equation}
\Phi=-GM_* \left(r^n+r_0^n\right)^{-1/n},
\label{gravpot}
\end{equation}
where $G$ is the gravitational constant, $M_*$ is the mass of the central object,
$r$ is the spherical radius, $r_0$ is the
softening radius, and $n=5$; tentatively, $r_0$ can be identified with the stellar radius.

\subsection{The initial state}
\label{Sec-initial}

Our initial state is the hydrostatic equilibrium obtained by
solving, for $h$, the vertical balance equation obtained from \Eq{momentumH},
\EQ
  -{\partial \over\partial z}(h+\Phi) + h{\partial s/c_p\over\partial z} = 0,
\label{VerticalBalance}
\EN
from large $z$ (where $h+\Phi=0$) down to $z=0$.
The initial density distribution $\varrho_0$ is then obtained using Eq.~(\ref{rrho});
it decreases monotonically with both $z$ and $\varpi$ in the equilibrium state.

The
initial rotation velocity, $u_{\varphi0}$, follows from the radial
balance equation,
\EQ
  - {u_{\varphi0}^2\over\varpi}
  = - {\partial\over\partial\varpi}(h+\Phi)
    + h{\partial s/c_p\over\partial\varpi}.
\label{uphi}
\EN
In the disc, $h=c_p T$ is small, so $u_{\varphi0}$ is close to the
Keplerian velocity, while the corona does not rotate initially,
and so is supported by the pressure gradient.

As a rough estimate, the value of $h$ in the midplane of the
disc is
\EQ
h_{\rm disc}\approx\beta\ h_{\rm corona}
\equiv\beta GM_*/\varpi,
\label{hdisc}
\EN
as can be seen by integrating
\Eq{VerticalBalance}, ignoring the $\partial\Phi/\partial z$ term.
Thus, the initial toroidal velocity in the midplane can be obtained from \Eq{uphi}
using \Eq{hdisc} and recalling that $\partial s/\partial\varpi=0$
in the midplane, which gives
\EQ
u_{\varphi0}\approx\sqrt{(1-\beta)GM_*/\varpi}=\sqrt{1-\beta}\vec{\uu}_{\rm K}.
\EN
For $\beta=0.1$, for example, the toroidal velocity is within 5\%
of the Keplerian velocity.


\subsection{Dimensionless variables and choice of parameters} \label{DVCP}

Our model is scale invariant and can therefore be applied to various
astrophysical objects.
We consider here the range of parameters typical of protostellar discs, for which a typical
surface density is $\Sigma_0\approx1\g\cm^{-2}$.
A typical coronal sound speed is $c_{\rm s0} \approx 10^2\km\s^{-1}$,
which corresponds to a temperature of $T \approx 4\times10^5\K$.
This allows us to fix relevant units as follows:
\EQ
[\Sigma]=\Sigma_0,\ \;
[u]=c_{\rm s0},\ \;
[x]=GM_* c_{\rm s0}^{-2},\ \;
[s]=c_p.
\label{nondim}
\EN
With $M_*=1\,M_\odot$ we have $[x] \approx 0.1\AU$, and $[t]=[x]/[u] \approx 1.5\,{\rm day}$.
The unit for density is $[\varrho]=[\Sigma]/[x] \approx 7\times10^{-13}\g\cm^{-3}$,
the unit for the mass accretion rate is
$[\dot{M}]=[\Sigma] [u] [x] \approx 2\times10^{-7}M_\odot\yr^{-1}$,
and the unit for the magnetic field is
$[B]=[u] (\mu_0[\varrho])^{1/2} \approx 30\G$.

Since $[h]=[u]^2$,
the dimensionless value $h=1$ corresponds
to $10^{14}\cm^2\s^{-2}$.
With a mean specific weight $\mu=0.6$, the universal gas constant
${\cal R}=8.3\times10^7\cm^2\s^{-2}\K^{-1}$ and $\gamma=5/3$, we have
\EQ
c_p={\gamma\over\gamma-1}{{\cal R}\over\mu}
\approx3.5\times10^8\cm^2\s^{-2}\K^{-1}.
\EN
Therefore, $h=1$ corresponds to
$T=[u]^2/[s]=[h]/c_p\approx3\times10^5\K$.
Using $h=-\Phi$ in the corona, this corresponds to a
temperature of $3\times10^5\K$ at $r=[x]\equiv0.1\AU$.

We choose $\beta$ between 0.1 and 0.005, corresponding to
a typical disc temperature (in the model) of $3\times10^4\K$ to
$1.5\times10^3\K$; see \Eq{hdisc};
real
protostellar discs have typical temperatures of about a few
thousand Kelvin (see Sect.~\ref{Sec-cool-hot}).

In our models, we use
$\varpi_0=1.5$ for the disc outer radius,
$z_0=0.15$ for the disc semi-thickness and
$r_0=0.05$ for the softening (stellar) radius.
The disc aspect ratio is $z_0/\varpi_0=0.1$.
Note that $r_0=0.05$ corresponds
to $7\times10^{10}\cm$, i.e.\ one solar radius.
Therefore, we shall not reduce it much below this physically meaningful value.
Note, however, that smaller values of $r_0$ would result in faster
outflows (Ouyed \& Pudritz 1999).

Furthermore, $c_\nu^{\rm adv}=0.02$ and $c_\nu^{\rm shock}=1.2$.
We vary the value of $\alpha_{\rm SS}$ between 0.003 and 0.007.

The mean-field dynamo is characterized by the parameters
$|\alpha_0|=0.3$, $v_0=\cs$,
$\eta_{{\rm t}0}=10^{-3}$ and $\eta_0=5\times10^{-4}$.
The total magnetic diffusivity in the disc is therefore
$\eta_{\rm T0}=\eta_{\rm t0}+\eta_0 = 0.0015$.
In terms of the usual Shakura--Sunyaev viscosity parameter,
this corresponds to
\EQ
\alpha_{\rm SS}^{(\eta)}\equiv
{\eta_{\rm T0}\over c_{\rm s,disc}z_0}
\approx0.01\left({\varpi\over\beta}\right)^{1/2}
\left({\eta_{\rm T0}\over0.0015}\right)
\left({z_0\over0.15}\right)^{-1}\!\!\!,
\label{alphaSS_eta}
\EN
where we have used $c_{\rm s,disc}^2\approx\beta c_{\rm s,corona}^2$
[cf.\ Eqs~(\ref{rrho}) and (\ref{hdisc})] and
$c_{\rm s,corona} \approx \varpi^{-1/2}$.

In terms of the usual nondimensional dynamo parameters we have
\EQ
|C_\alpha|=|\alpha_0| z_0/\eta_{\rm T0}=30
\EN
and, for Keplerian rotation,
\EQ
C_\omega=(\varpi\partial\Omega/\partial\varpi)z_0^2/\eta_{\rm T0}
=-22.5\varpi^{-3/2},
\EN
so that the dynamo number is given by
\EQ
|{\cal D}|=|C_\alpha C_\omega|=675\varpi^{-3/2}.
\label{CalpCom}
\EN
Note that the value of the dynamo number expected for accretion discs is given by
\EQ
|{\cal D}|=\frac{ |\alpha_0\,\varpi\partial\Omega/\partial\varpi|\,z_0^3}{\eta_{\rm T0}^2}
\la{3\over2}\left(\alpha_{\rm SS}^{(\eta)}\right)^{-2}
\label{calD}
\EN
for $\eta_{\rm T0}=\alpha_{\rm SS}^{(\eta)}c_{\rm s,disc}z_0$,
$|\alpha_0|\la c_{\rm s,disc}$ and $c_{\rm s,disc}=\Omega z_0$.
As follows from Eqs~(\ref{alphaSS_eta})--(\ref{CalpCom}),
for Keplerian rotation,
\EQ
{2\over 3} |{\cal D}| \left(\alpha_{\rm SS}^{(\eta)}\right)^2
= |\alpha_0| z_0 \beta^{-1} \varpi^{-1/2}.
\label{la1}
\EN
[Note that this expression is independent of $\eta_{\rm T0}$.]
For our choice of parameters of $|\alpha_0|=0.3$ and $z_0=0.15$, expression (\ref{la1})
is $\la 1$
if $\varpi \ga 0.002\ \beta^{-2} = 0.2$ for $\beta=0.1$, which corresponds to
the truncation radius of the $\alpha$-effect. Therefore,
our choice of parameters is consistent with the constraint (\ref{calD}).

We take $\gamma=5/3$, $\tau_{\rm disc}=0.1$,
and consider two values of $\tau_{\rm star}$. For
$\tau_{\rm star}\to\infty$, the mass sink at the central object
is suppressed, whereas $\tau_{\rm star}\to0$ implies instantaneous
accretion of any extra matter (relative to the hydrostatic equilibrium)
by the central object. A realistic
lower limit for $\tau_{\rm star}$ can be estimated
as $\tau_{\rm star}\ga r_0/v_{\rm ff}\approx0.008$, where $v_{\rm ff}$ is the free fall
velocity (given by $\half v_{\rm ff}^2=GM_*/r_0$ in dimensional quantities).
In most cases we used
$\tau_{\rm star}=0.01$, but we also tried an even smaller value of
$\tau_{\rm star}=0.005$ and obtained very similar results.
A finite value of $\tau_{\rm star}$ implies that matter is not
instantaneously absorbed by the sink. Therefore, some matter can leave
the sink if it moves so fast that its residence time in the sink
is shorter than $\tau_{\rm star}$. As can be seen below, a small (negligible)
fraction of mass does indeed escape from the sink.

Computations have been carried out in domains ranging from
$(\varpi,z)\in[0,2]\times[-1,1]$ to $[0,8]\times[-2,30]$, but
the results are hardly different
in the overlapping parts of the domains.
In our standard computational box,
$\delta \varpi=\delta z=0.01$
and in the case of the larger computational domain
$\delta \varpi=\delta z=0.02$.

\subsection{Numerical method and boundary conditions}

We use third order Runge--Kutta time-stepping and a sixth order
finite-difference scheme in space. Details and test calculations are discussed by
Brandenburg (2001).

On the outer boundaries, the induction equation is evolved using
one-sided derivatives (open boundary conditions).
The normal velocity component has
zero derivative normal to the boundary, but the velocity
is required to be always directed outwards.
The tangential velocity
components and potential enthalpy on the boundaries are similarly obtained
from the next two interior points away from the boundary.
Tests show that the presence of the boundaries does not affect the flow
inside the computational domain. Regularity
conditions are adopted on the axis where
the $\varpi$ and $\varphi$ components of vectorial quantities vanish,
whereas scalar variables and the $z$ components of vectorial quantities
have vanishing radial derivative.

\begin{figure}[t!]
\centerline{%
  \includegraphics[width=8.5cm]{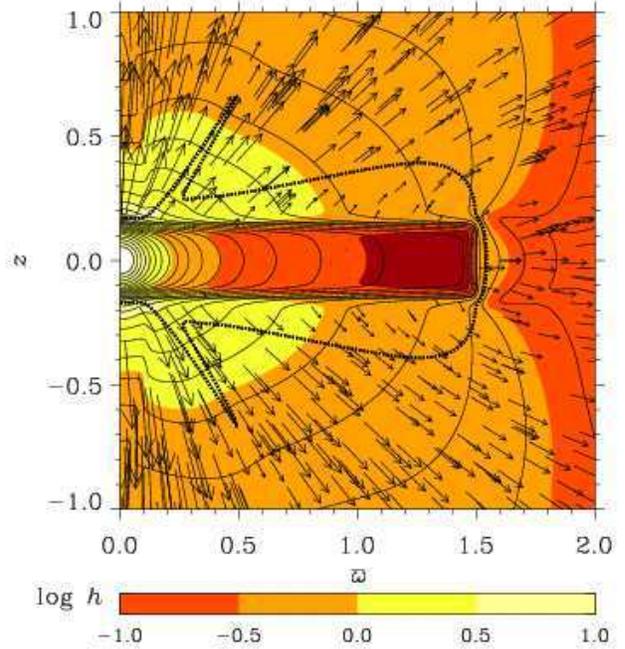}}
\caption[]{A nonmagnetic model without mass sink at the centre: velocity vectors and
logarithmically spaced density contours
(from 0.008 to 6000, separation factor of 1.6) superimposed on a
grey scale representation of $\log_{10}h$ at time $t=223$.
Specific enthalpy $h$ is directly proportional to temperature $T$, and
$\log_{10}h=(-1,0,1)$ corresponds to
$T\approx (3{\times}10^{4},3{\times}10^{5},3{\times}10^{6})\,\mbox{K}$.
Note hot gas near the central object and in the near corona, and cooler gas
in the disc and the far corona. The dashed line shows the sonic surface where the
poloidal velocity equals $\cs$.
The disc boundary is shown with a thin black line; $\beta=0.1$.
}\label{Flito0}\end{figure}

\begin{figure}[t!]
\centerline{%
  \includegraphics[width=8.5cm]{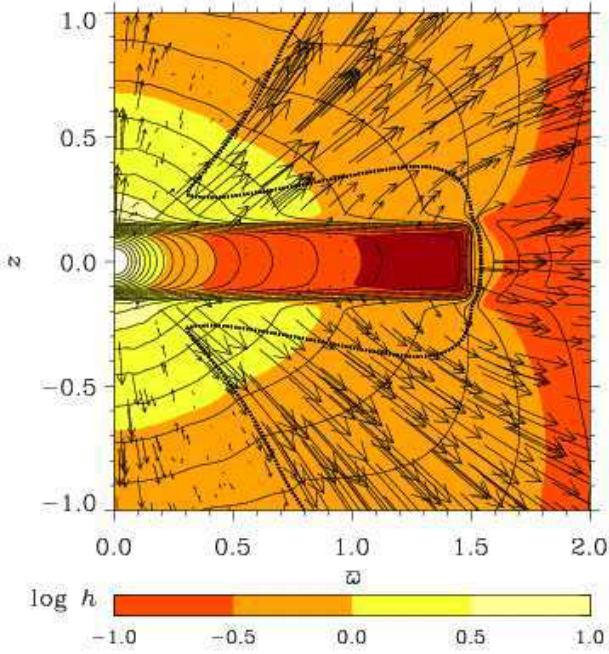}}
\caption[]{
As in Fig.~\ref{Flito0}, but
with a mass sink at the centre, $\tau_{\rm star}=0.01$, at time $t=312$.
The outflow speed
near the axis is strongly reduced in comparison to that in the model without mass sink
shown in Fig.~\ref{Flito0}.
}
\label{FRun9}
\end{figure}

\section{Results} \label{RS}
In this section we discuss a range of models of increasing degree of complexity.
We first consider in Sect.~\ref{NMO} the simplest model that contains neither magnetic field nor mass
sink at the centre to show how the pressure gradient resulting from the (fixed) entropy
distribution drives a disc outflow. It is further shown that the outflow is
significantly reduced if accretion onto the central object is allowed, but restored again if
the disc can generate a large-scale magnetic field (Sect.~\ref{MO}). Having thus demonstrated the
importance of the large-scale magnetic field in our model, we discuss in Sect.~\ref{MFS} its structure,
largely controlled by the dynamo action but affected by the outflow. Model parameters used
in these sections are not necessarily realistic as we aim to illustrate the general physical nature
of our solutions. We present a physically motivated model in Sect.~\ref{FinalModel} where the set of
model parameters is close to that of a standard accretion disc around a protostellar object.
The physical nature of our solutions is discussed in Sects.~\ref{M-E-loss}, \ref {MWA} and \ref{Lagr}.

\subsection{Nonmagnetic outflows}     \label{NMO}

We illustrate in  Fig.~\ref{Flito0} results obtained for a model
without any magnetic fields and without mass sink.
A strong outflow develops even in this case,
which is driven mostly by the vertical pressure gradient in
the transition layer between the disc and the corona, in particular by the term
$T\vec{\nabla} s$ in Eq.~(\ref{Tgrads_in_pgrad}).
The gain in velocity is controlled by the total
specific entropy difference between the disc and the corona, but not by
the thickness $d$ of the transition layer in the disc profile
(\ref{xi}).

The flow is fastest along the rotation axis and within a cone of polar
angle of about $30^\circ$,
where the terminal velocity $u_z\approx3$ is reached.
The conical shape of the outflow is partly due to obstruction from
the dense disc, making it easier for matter to leave the
disc near the axis. Both temperature and
density in nonmagnetic runs without mass sink are reduced very
close to the axis
where the flow speed is highest.

The general flow pattern is sensitive to whether or not matter can accrete
onto the central object. We show in Fig.~\ref{FRun9} results for the
same model as in Fig.~\ref{Flito0}, but with a mass sink given by Eq.~(\ref{qminus})
with $\tau_{\rm star}=0.01$.
This can be compared with earlier work on thermally driven winds by
Fukue (1989) who also considered polytropic outflows, but the disc
was treated as a boundary condition. In Fukue (1989), outflows are driven when the
injected energy is above a critical value. The origin of this
energy injection may be a hot corona. The critical surface in
his model (see the lower panel of his Fig.~2) is quite similar
to that found in our simulation (our Fig.~\ref{FRun9}), although
our opening angle was found to be larger than in Fukue's (1989) model.
[Below we show, albeit with magnetic fields, that smaller values
of $\beta$ do result in smaller opening angles, see Fig.~\ref{FRun7_0.02},
which would then be compatible with the result of Fukue (1989).]

As could be expected, the mass sink hampers the outflow in the cone (but not at
$\varpi \ga 0.5$).
The flow remains very similar to that of Fig.~\ref{FRun9} when $\tau_{\rm star}$ is reduced
to $0.005$.
Thus, the nonmagnetic outflows are very sensitive to the presence of the
central sink.
As we show now, magnetized outflows are different in this respect.

\subsection{Magnetized outflows} \label{MO}

In this section we discuss results obtained with magnetic fields,
first without mass sink at the centre
and then including a sink. We show that the effects of the sink are
significantly weaker than in the nonmagnetic case.

\begin{figure}[t!]
\centerline{%
  \includegraphics[width=8.5cm]{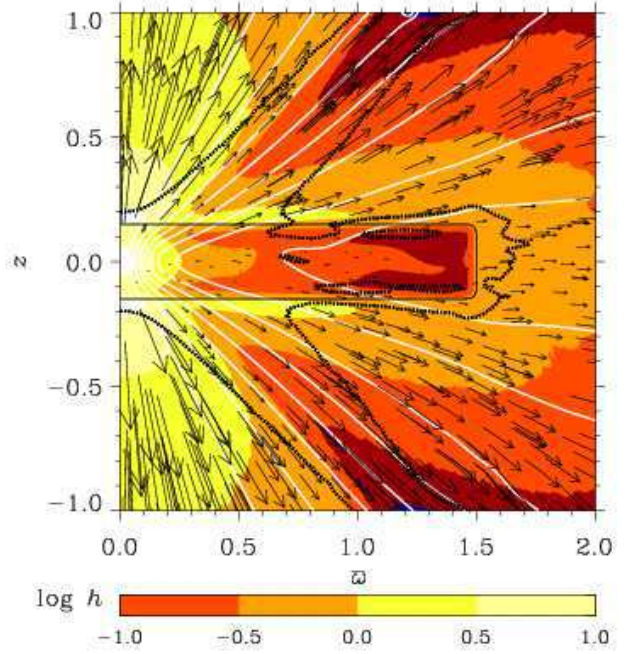}}
\caption[]{
A magnetic model without mass sink at the centre:
velocity vectors and poloidal magnetic field lines (white)
superimposed on a grey scale representation of
temperature (in terms of $h$) for a run with
$\alpha_0=-0.3$ (resulting in roughly dipolar magnetic symmetry) at
time $t=269$.
In contrast to the nonmagnetic model of Fig.~\protect\ref{Flito0},
a conical shell has developed that is cooler and less dense than its exterior.
The conical shell intersects $z=\pm1$ at around $\varpi\approx 1.2$.
The dashed line shows the
surface
where the poloidal velocity equals
$(\cs^2+v_{\rm A,pol}^2)^{1/2}$,
with $v_{\rm A,pol}$ the Alfv\'en speed from the poloidal magnetic field
(fast magnetosonic surface with respect to the poloidal field).
The disc boundary is shown with a thin black line; $\beta=0.1$.
}\label{Flito1}\end{figure}

\begin{figure}[t!]
\centerline{%
  \includegraphics[width=8.5cm]{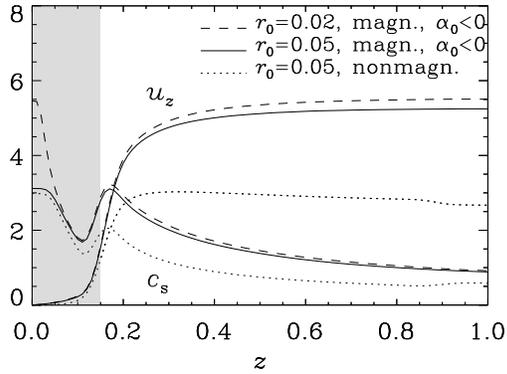}}
\caption[]{
Profiles of vertical velocity and sound speed along
the axis, $\varpi=0$, for the
nonmagnetic model of Fig.~\protect\ref{Flito0} with smoothing radius $r_0=0.05$
(dotted) and two runs with magnetic
field, with $r_0=0.05$ as in Fig.~\protect\ref{Flito1} (solid) and $r_0=0.02$ (dashed, $t=162$).
The shaded area marks the location of the disc. Terminal
wind speeds are reached after approximately three disc heights.
The presence of a magnetic field and a deeper potential well
(smaller $r_0$) both result in faster outflows.
The models shown here have no mass sink at the centre and $\beta=0.1$.
}\label{Fpvz}\end{figure}

\begin{figure*}[t!]
\centerline{%
  \includegraphics[width=18cm]{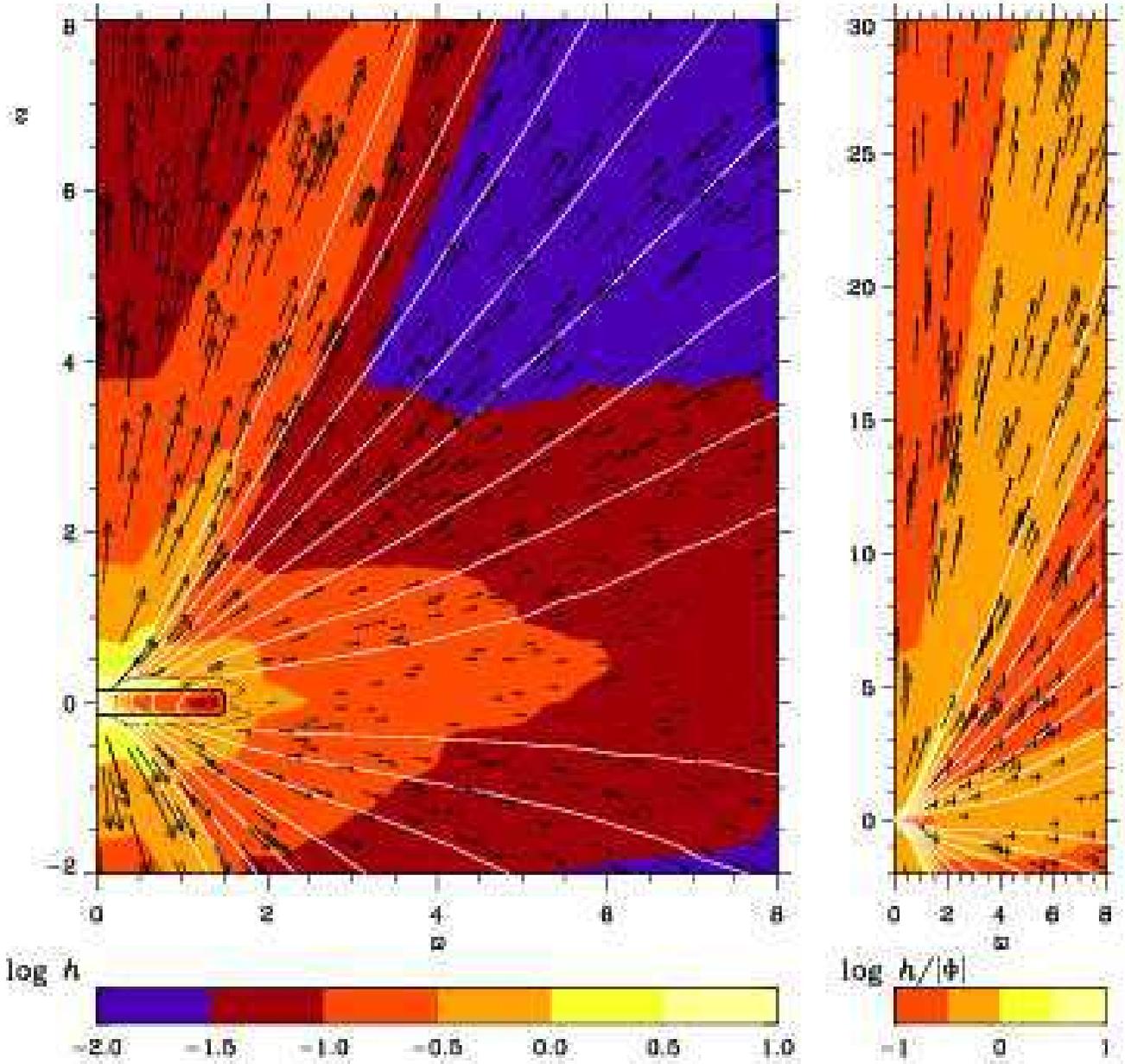}}
\caption[]{Results for a larger domain, $[0,8]\times[-2,30]$, with the
same parameters as in Fig.~\protect\ref{Flito1}, averaged over times
$t=200 \dots 230$, $\beta=0.1$.
Left panel: velocity vectors, poloidal magnetic field lines
and grey scale representation of $h$
in the inner part of the domain.
Right panel: velocity vectors, poloidal magnetic field lines and
normalized specific enthalpy $h/|\Phi|$ in the full domain.
}\label{FRRRRun3c}\end{figure*}

A magnetized outflow without the central mass sink, shown in Fig.~\ref{Flito1}, is similar
to that in Fig.~\ref{Flito0}, but is denser and hotter near the axis,
and the high speed cone has a somewhat larger opening angle.
In addition, the outflow becomes more structured, with a well pronounced
conical shell where temperature and density are smaller than elsewhere
(the conical shell reaches
$z=\pm1$ at $\varpi\approx1.2$ in Fig.~\ref{Flito1}).
Here and in some of the following figures we also show the fast
magnetosonic surface with respect to the poloidal field.
In Sect.~\ref{MWA} we show that this surface is close
to the fast magnetosonic surface.
As shown in  Fig.~\ref{Fpvz}, the outflow becomes
faster inside the cone ($u_z \approx 5$ on the axis).
As expected, we find that
deeper potential wells, i.e.\ smaller values of $r_0$ in Eqs~(\ref{xistar}) and (\ref{gravpot}),
result in even faster flows 
and in larger opening angles.

Our results are insensitive to the size and symmetry of the computational domain: we
illustrate this in Fig.~\ref{FRRRRun3c} with a larger
domain of the size $[0,8]\times[-2,30]$.
The disc midplane in this run is located asymmetrically in $z$
in order to verify that the (approximate) symmetry of the solutions
is not imposed by the symmetry of the computational domain. Figure~\ref{FRRRRun3c} confirms that our results
are not affected by what happens near the computational boundaries.

\begin{figure}[t!]
\centerline{%
  \includegraphics[width=8.5cm]{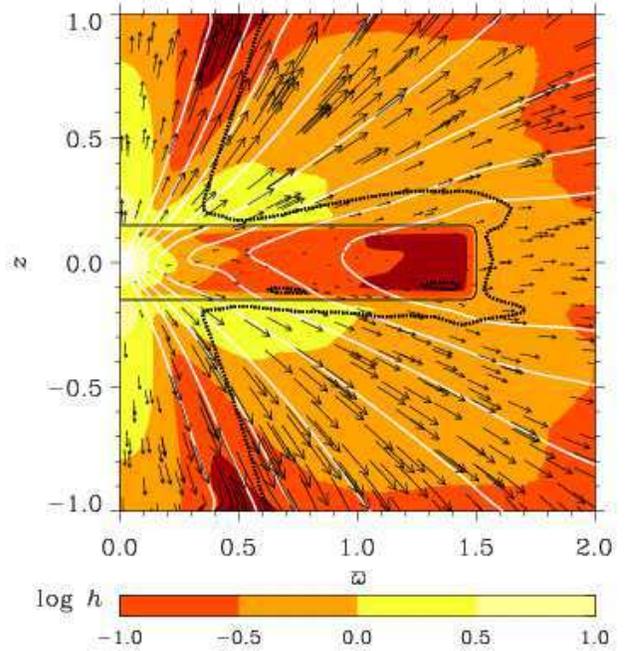}}
\caption[]{
As in Fig.~\protect\ref{Flito1}, but with a mass sink at the centre,
$\tau_{\rm star}=0.01$.
The outflow speed near the axis and the opening angle of the conical shell
(now reaching $z=\pm1$ at $\varpi\approx0.5$)
are reduced in comparison to that in
the model without mass sink shown in Fig.~\ref{Flito1},
but the general structure of the outflow is little affected.
Averaged over times $t=130 \dots 140$, $\beta=0.1$.
[Reference model.]
}\label{FRun7}\end{figure}

\begin{figure}[t!]
\centerline{%
  \includegraphics[width=8.5cm]{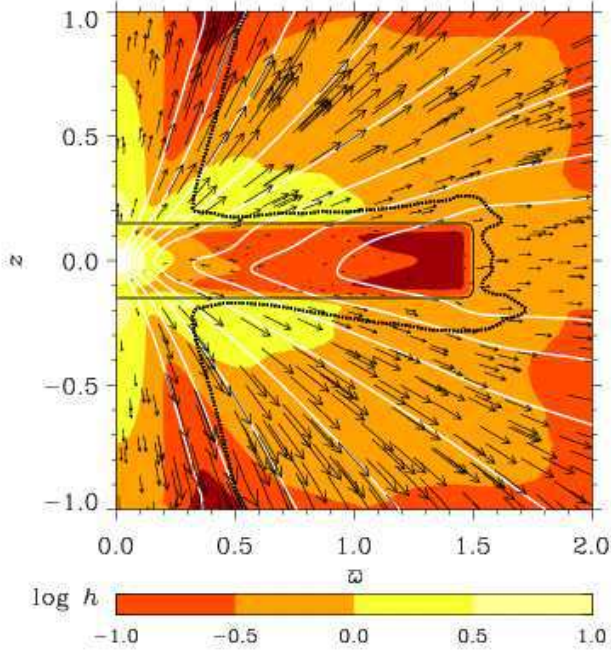}}
\caption[]{
As in Fig.~\protect\ref{FRun7}, but with a stronger mass sink at the centre,
$\tau_{\rm star}=0.005$.
The flow pattern is very similar to that of Fig.~\protect\ref{FRun7}
where the time scale of the sink is twice as large.
Averaged over times $t=122 \dots 132$, $\beta=0.1$.
}\label{FRun8}\end{figure}

Unlike the nonmagnetized system, the magnetized outflow changes only comparatively
little when the mass sink is introduced at the centre. We show in Fig.~\ref{FRun7}
the results with a sink ($\tau_{\rm star}=0.01$) and otherwise the same
parameters as in Fig.~\ref{Flito1}. As could be expected, the sink leads to a
reduction in the outflow speed near the axis; the flow in the high speed cone becomes slower.
But apart from that, the most
important effects of the sink are the enhancement of the conical structure of the
outflow and the smaller opening angle of the conical shell.
A decrease in $\tau_{\rm star}$ by a factor of 2 to $0.005$ has very little effect,
as illustrated in Figs.~\ref{FRun8} and \ref{Fpvaccall}.

\begin{figure}[t!]
\centerline{%
  \includegraphics[width=8.5cm]{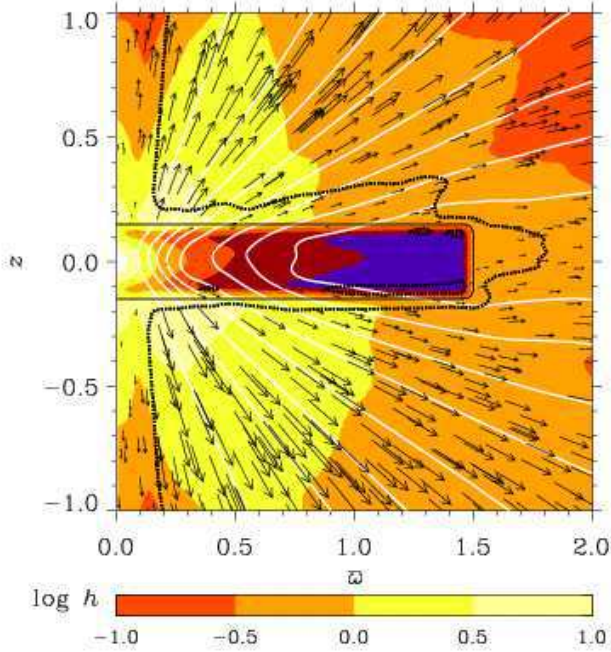}}
\caption[]{
As in Fig.~\protect\ref{FRun7}, but with a larger entropy contrast,
$\beta=0.02$.
The opening angle of the conical shell is reduced in comparison to that of
Figs.~\protect\ref{FRun7} and \protect\ref{FRun8} where the entropy contrast
is smaller (the shell crosses $z=\pm1$ at $\varpi\approx0.15$).
Averaged over times $t=140...240$, $\tau_{\rm star}=0.01$.
}\label{FRun7_0.02}\end{figure}

\begin{figure}[t!]
\centerline{%
  \includegraphics[width=8.5cm]{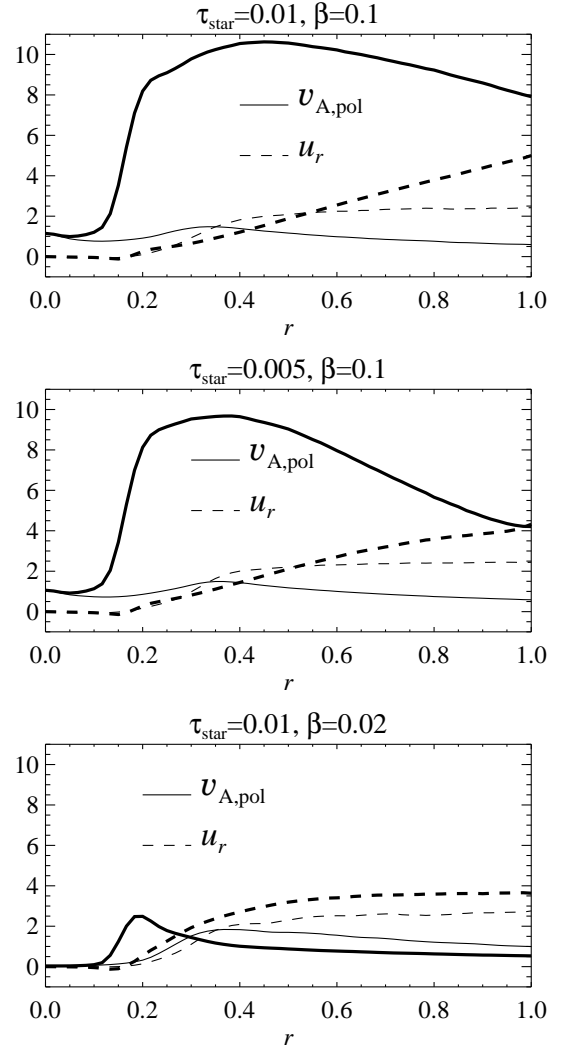}}
\caption[]{
Spherical radial velocity component, $u_r$ (dashed), and poloidal
Alfv\'en speed, $v_{\rm A,pol}$ (solid), as functions of spherical
radius at polar angles $\vartheta = 30^\circ$ (thick lines) and
$\vartheta = 60^\circ$ (thin lines) for the models of
Figs.~\protect\ref{FRun7} (top panel),
\protect\ref{FRun8} (middle panel) and
\protect\ref{FRun7_0.02} (bottom panel).
}\label{Fpvaccall}\end{figure}

Increasing the entropy contrast (while keeping the specific entropy unchanged in the corona)
reduces the opening angle of the conical shell.
Pressure driving is obviously more important in this case, as compared
to magneto-centrifugal driving (see Sect.~\ref{MWA}).
A model with $\beta=0.02$ (corresponding to a density and inverse temperature
contrast of about 50:1 between the disc and the corona) is shown in
Fig.~\ref{FRun7_0.02}.
At $\vartheta=60^\circ$, the radial velocity $u_r$ is slightly
enhanced relative to the case $\beta=0.1$ (contrast 10:1);
see Fig.~\ref{Fpvaccall}.
At $\vartheta=30^\circ$, on the other hand, the flow with the larger entropy contrast
reaches the Alfv\'en point close to the disc (at $r\approx0.27$ as opposed to $r \ga 1$
in the other case), which leads to a smaller terminal velocity.

We conclude that the general structure of the
{\em magnetized\/} flow and its typical parameters
remain largely unaffected by the sink, provided its efficiency $\tau_{\rm star}^{-1}$ does not exceed a certain threshold.
It is plausibly the build-up of magnetic pressure at the centre that shields the central
object to make the central accretion inefficient. This shielding would be
even stronger if we included a magnetosphere of the central object. We discuss in
Sect.~\ref{LS} the dependence of our solution on the geometrical size of the sink and
show that the general structure of the outflow persists as long as the size of the sink does
not exceed the disk thickness.

\subsection{Magnetic field structure} \label{MFS}

The dynamo in most of our models has $\alpha_0<0$, consistent with results from simulations of
disc turbulence driven by the magneto-rotational instability
(Brandenburg et al.\ 1995; Ziegler \& R\"udiger 2000). The
resulting field symmetry is roughly dipolar,
which seems to be typical of $\alpha\Omega$ disc
dynamos with $\alpha_0<0$ in a conducting corona (e.g.,\ Brandenburg et al.\
1990). We note that the dominant parity of the magnetic field is sensitive
to the magnetic diffusivity in the corona:
a quadrupolar oscillatory magnetic field dominates for $\alpha_0<0$
if the disc is surrounded by vacuum (Stepinski \& Levy 1988).

For $\alpha_0<0$, the critical value of $\alpha_0$ for dynamo action is
about 0.2, which is a factor of about 50 larger than without
outflows.
Our dynamo is then only less than twice supercritical. A
survey of the dynamo regimes for similar models is given by
Bardou et al.\ (2001).

\begin{figure}[t!]
\centerline{%
  \includegraphics[width=8.5cm]{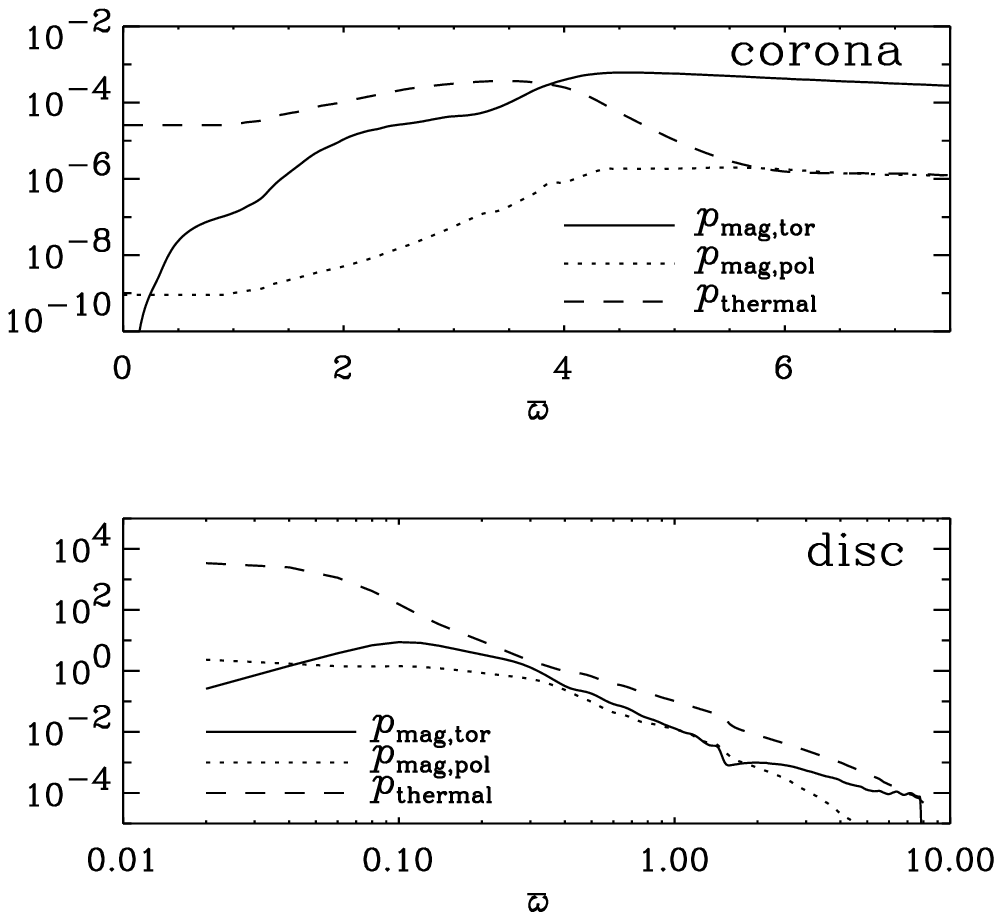}}
\caption[]{
Radial profiles of magnetic pressure from the toroidal (solid) and poloidal (dotted)
magnetic fields and thermal pressure (dashed)  for the model of Fig.~\protect\ref{FRRRRun3c}.
Shown are the averages over the disc volume (lower panel) and
over a region of the same size around $z=8$ in the corona (upper panel). $\beta=0.1$.
}\label{Fpbint}\end{figure}

The initial magnetic field (poloidal, mixed parity) is weak
[$p_{\rm mag}\equiv\BB^2/(2\mu_0)\approx10^{-5}$], cf.\ Fig.~\ref{Fpbint}
for comparison with the gas pressure],
but the dynamo soon amplifies the field in the
disc to $p_{\rm mag,tor}\approx10$, and then supplies it to the corona. As a result, the corona is
filled with a predominantly azimuthal field with $p_{\rm mag,tor}/p\approx100$ at larger radii; see
Fig.~\ref{Fpbint}. We note, however, that the flow in the corona varies significantly in
both space and time\footnote{See movie at
\url{http://www.nordita.dk/~brandenb/movies/outflow}
}. Magnetic pressure due to the toroidal field $B_\varphi$, $p_{\rm mag,tor}$, exceeds
gas pressure in the corona outside the inner cone
and confines the outflow in the conical shell. The main mechanisms producing
$B_\varphi$ in the corona are advection by the wind and magnetic buoyancy (cf.\ Moss
et al.\ 1999). Magnetic diffusion and stretching of the poloidal field by
vertical shear play a relatively unimportant r\^ole.

The field in the inner parts of the disc is dominated by the toroidal component;
$|B_\varphi/B_z|\approx3$ at $\varpi\la0.5$; this ratio is larger
in the corona at all $\varpi$.
However, as shown in  Fig.~\ref{Fpbint}, this ratio is closer to unity at larger radii
in the disc.

\begin{figure}[t!]
\centerline{%
  \includegraphics[width=8.5cm]{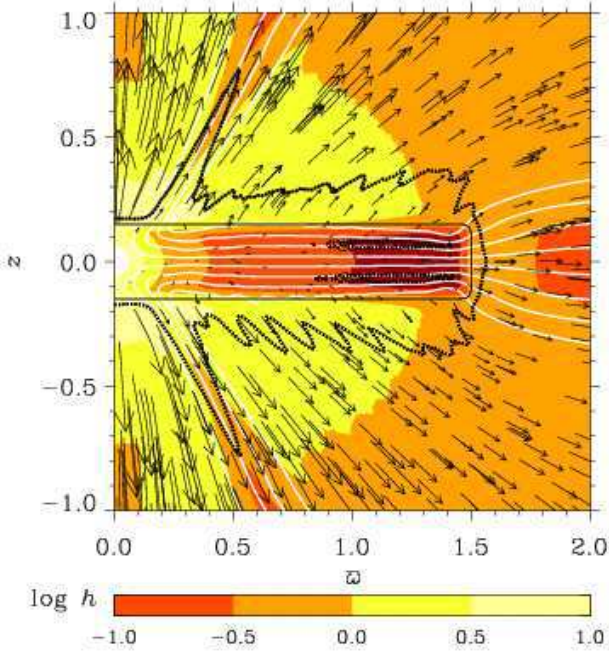}}
\caption[]{
As in Fig.~\ref{Flito1}, but with $\alpha_0=+0.3$, at time  $t=132$.
The magnetic field geometry is now mostly quadrupolar because $\alpha_0>0$.
}\label{Flito2}\end{figure}

As expected, $\alpha_0>0$ results in mostly quadrupolar fields (e.g.,\ Ruzmaikin
et al.\ 1988). As shown in Fig.~\ref{Flito2}, the magnetic field in the corona
is now mainly restricted to a narrow conical shell that crosses $z=\pm1$ at $\varpi\approx0.6$.
Comparing this figure with the results
obtained with dipolar magnetic fields (Fig.~\ref{Flito1}),
one sees that the quadrupolar field has
a weaker effect on the outflow than the dipolar field; the conical shell is less pronounced.
However, the structures within the inner cone are qualitatively similar to each other.

The magnitude and distribution of $\alpha$ in Eq.~(\ref{alpha}) only weakly
affect magnetic field properties as far as the dynamo is saturated. For a saturated dynamo, the
field distribution in the dynamo region ($0.2<\varpi<1.5$, $|z|<0.15$)
roughly follows from the equipartition field, $B\simeq(\varrho\mu_0 v_0^2)^{1/2}$
with $v_0=c_{\rm s}$.
In other words, nonlinear states of disc dynamos are almost insensitive to the
detailed properties of $\alpha$ (e.g.,\ Beck et al.\ 1996; Ruzmaikin et al.\ 1988).

A discussion of disc dynamos with outflows, motivated by the present model, can
be found in Bardou et al.\ (2001). It is shown that the value of magnetic
diffusivity in the corona does not affect the dynamo solutions strongly.
Moreover, the outflow is fast enough to have the magnetic Reynolds number
in the corona larger than unity, which implies that ideal integrals
of motion are very nearly constant along field lines;
see Sect.~\ref{Lagr}. The most important
property is the sign of $\alpha$ as it controls the global symmetry of the
magnetic field.

\subsection{Mass and energy loss}
\label{M-E-loss}

The mass injection and loss rates
due to the source, sink and wind are defined as
\begin{equation}
\dot{M}_{\rm source} = \int q_\varrho^{\rm disc} \, \dd V,\quad
\dot{M}_{\rm sink} = \int q_\varrho^{\rm star} \, \dd V,
\label{mass_source+sink}
\end{equation}
and
\begin{equation}
\dot{M}_{\rm wind}=\oint\varrho \uu\cdot\dd\SSS,
\label{mass_source_surf}
\end{equation}
respectively, where
the integrals are taken over the full computational domain
or its boundary.
About $1/3$ of the mass released goes into the wind and the rest is accreted by the sink,
in the model with $\tau_{\rm star} = 0.01$ and $\beta=0.1$ of Fig.~\ref{FRun7}.
Reducing $\tau_{\rm star}$ by a factor of $2$ (as in the model of Fig.~\ref{FRun8}), only changes
the global accretion parameters by a negligible amount ($\la 10 \%$).

\begin{figure*}[t!]
\centerline{%
  \includegraphics[width=18cm]{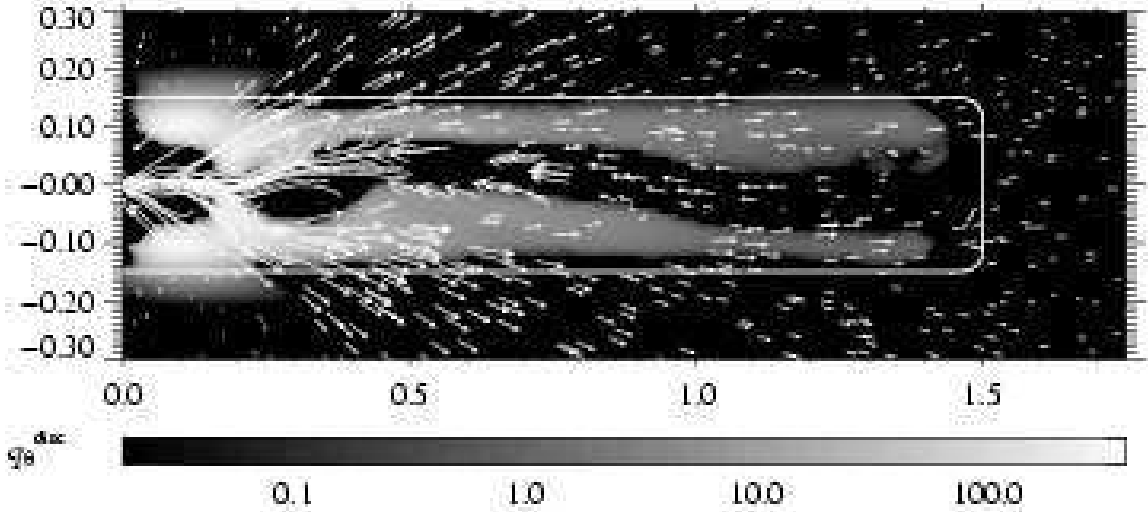}}
\caption[]{
Azimuthally integrated mass flux density, represented as a vector
$2\pi \varpi \varrho (u_\varpi, u_z)$, in the simulation of Fig.~\protect\ref{FRun7} with
a dipolar magnetic field and a mass sink at the centre. Shades of grey show
the distribution of the mass source in the disc, $q_\varrho^{\rm disc}$. The disc boundary is shown
with a white line.
}\label{stream}\end{figure*}

The mass loss rate in the wind fluctuates on a time
scale of 5 time units,
but remains constant on average at about $\dot M_{\rm wind}\approx3$, corresponding
to $6\times10^{-7}\,M_\odot\yr^{-1}$, in the models of Figs.~\ref{FRun7} and \ref{FRun8}.
The mass
in the disc, $M_{\rm disc}$, remains roughly constant.

The rate at which mass needs to
be replenished in the disc, $\dot M_{\rm source}/M_{\rm disc}$, is about 0.4.
This rate is not controlled by
the imposed response rate of the mass source, $\tau_{\rm disc}^{-1}$, which is 25 times larger. So,
the mass source adjusts itself to the disc evolution and does not directly
control the outflow.
We show in Fig.~\ref{stream} trajectories that start in and around the
mass injection region.
The spatial distribution of the mass replenishment rate $q_\varrho^{\rm disc}$
shown in Fig.~\ref{stream} indicates that
the mass is mainly injected close to the mass sink, and
$q_\varrho^{\rm disc}$ remains moderate in the outer parts of the disc.
(Note that the reduced effect of the mass sink in the magnetized flow
is due to magnetic shielding rather than to mass replenishment near
the sink -- see Sect.~\ref{MO}.)

\begin{figure}[t!]
\centerline{%
  \includegraphics[width=9cm]{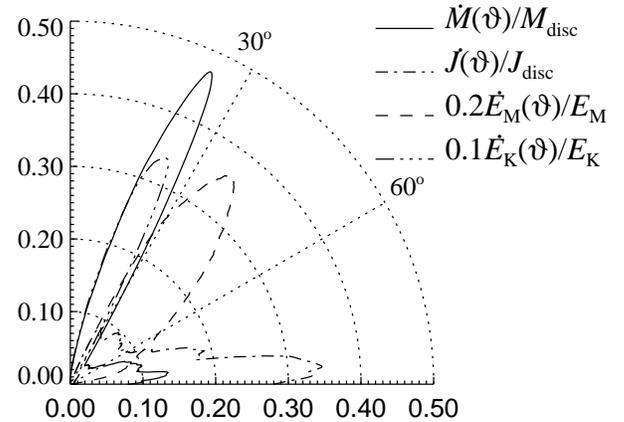}}
\caption{Dependence, on polar angle $\vartheta$, of azimuthally
  integrated radial mass flux density $\dot{M}(\vartheta)$ through a sphere $r=8$
  (solid, normalized by the disc mass $M_{\rm disc}\approx12$),
  azimuthally integrated radial angular momentum flux density $\dot{J}(\vartheta)$
  (dashed-dotted, normalized by the disc angular momentum $J_{\rm disc}\approx4.3$),
  azimuthally integrated radial Poynting flux density $\dot{E}_{\rm M}(\vartheta)$ (divided by 5, dashed,
  normalized by the magnetic energy in the disc $E_{\rm M}\approx0.6$),
  and azimuthally integrated radial kinetic energy flux density $\dot{E}_{\rm K}(\vartheta)$ (divided by 10,
  dash-3dots, normalized by the kinetic energy in the disc $E_{\rm K}\approx9.7$),
  for the model of Fig.~\protect\ref{FRRRRun3c}.
  The unit of all the quantities is $[t]^{-1}\,{\rm rad}^{-1}$.
  }
\label{Fpcircle_new}
\end{figure}

The angular structure of the outflow can be characterized by the following
quantities calculated for a particular spherical
radius, $r=8$,
for the model of Fig.~\ref{FRRRRun3c}: the azimuthally integrated normalized
radial mass flux density, $\dot{M}(\vartheta)/M_{\rm disc}$, where
\EQ
\dot{M}(\vartheta)=2\pi r^2\varrho u_r\sin\vartheta,\quad
M_{\rm disc}=\int_{\rm disc}\rho\,\dd V,
\EN
the azimuthally integrated normalized radial angular momentum flux density,
$\dot{J}(\vartheta)/J_{\rm disc}$, where
\EQ
\dot{J}(\vartheta)=2\pi r^2\varrho \varpi u_\varphi u_r\sin\vartheta,\quad
J_{\rm disc}=\int_{\rm disc}\rho \varpi u_\varphi\,\dd V,
\EN
the azimuthally integrated normalized radial magnetic energy flux (Poynting flux) density,
$\dot{E}_{\rm M}(\vartheta)/E_{\rm M}$, where
\EQ
\dot{E}_{\rm M}(\vartheta)=2\pi r^2{(\EE\times\BB)_r\over\mu_0}\sin\vartheta,\quad
E_{\rm M}=\int_{\rm disc}{\BB^2\over2\mu_0}\,\dd V,
\EN
and the azimuthally integrated normalized radial kinetic energy flux density,
$\dot{E}_{\rm K}(\vartheta)/E_{\rm K}$, where
\EQ
\dot{E}_{\rm K}(\vartheta)=2\pi r^2(\half\varrho\uu^2 u_r)\sin\vartheta,\quad
E_{\rm K}=\int_{\rm disc}\half\rho\uu^2\,\dd V.
\EN
Here, $M_{\rm disc},\ J_{\rm disc},\ E_{\rm M}$ and $E_{\rm K}$ are the mass,
angular momentum, and magnetic
and kinetic energies in the disc.
A polar diagram showing these distributions
is presented in Fig.~\ref{Fpcircle_new}
for the model of Fig.~\ref{FRRRRun3c}. Note that this is a model without
central mass sink where the flow in the hot, dense cone around the axis is fast.
The fast flow in the hot, dense
cone carries most of the mass
and kinetic energy.
A significant part of angular momentum is carried away in the
 disc plane whilst the magnetic field is ejected at intermediate angles,
 where the conical shell is located.

The radial kinetic and magnetic energy flux densities, integrated over the whole sphere,
are
$L_{\rm K}\equiv\int_0^\pi\dot{E}_{\rm K}(\vartheta)\,\dd\vartheta\approx0.54\,\dot{M}_{\rm wind}c_*^2$ and
$L_{\rm M}\equiv\int_0^\pi\dot{E}_{\rm M}(\vartheta)\,\dd\vartheta\approx0.03\,\dot{M}_{\rm wind}c_*^2$,
respectively, where $c_*\approx2.8$ is
the fast magnetosonic speed (with respect to the poloidal field)
at the critical surface (where $u_z=c_*$) on the axis.
Thus, $\dot{M}_{\rm wind}c_*^2$ can be taken as a good indicator of the kinetic energy
loss, and the magnetic energy loss into the exterior is about 3\% of this value.
These surface-integrated flux densities (or luminosities) are, as expected,
roughly independent of the distance from the central object.

\begin{figure}[t!]
\centerline{%
  \includegraphics[width=8.5cm]{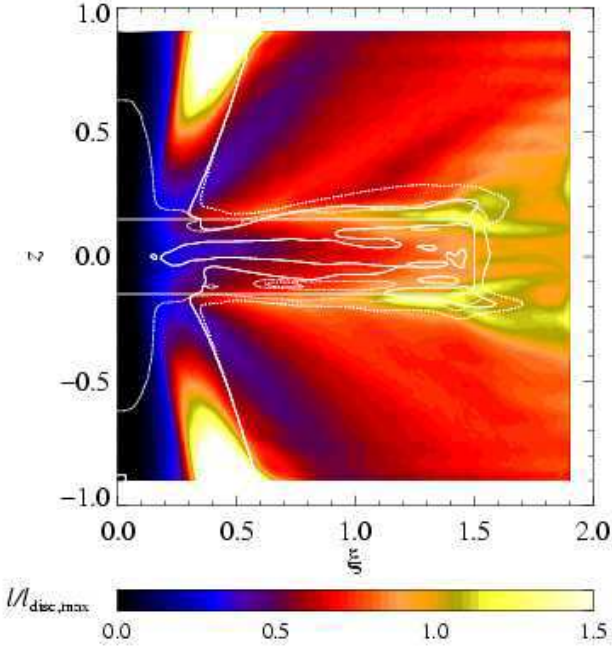}}
\caption[]{
Angular momentum (normalized by the maximum angular momentum in the disc)
for the model with $\tau_{\rm star} = 0.01$ and $\beta=0.1$, shown in
Fig.~\ref{FRun7}. The maximum value is $l/l_{\rm disc,max} \approx 2.5$.
The dashed line shows the
fast magnetosonic surface with respect to the poloidal field (cf.\ Fig.~\ref{Flito1}),
the solid line the Alfv\'en surface where the poloidal velocity equals
the poloidal Alfv\'en speed, and the dotted line the sonic surface.
Averaged over times $t=130...140$.
}\label{FRun7new01pangmom}\end{figure}

\subsection{Mechanisms of wind acceleration} \label{MWA}

The magnetized outflows in our models with central mass sink have a
well-pronounced structure, with a fast, cool and
low-density flow in a conical shell, and a slower, hotter and denser flow
near the axis and in the outer parts of the domain.
Without central mass sink, there is a high speed, hot and dense cone
around the axis.

\begin{figure}[t!]
\centerline{%
  \includegraphics[width=8.5cm]{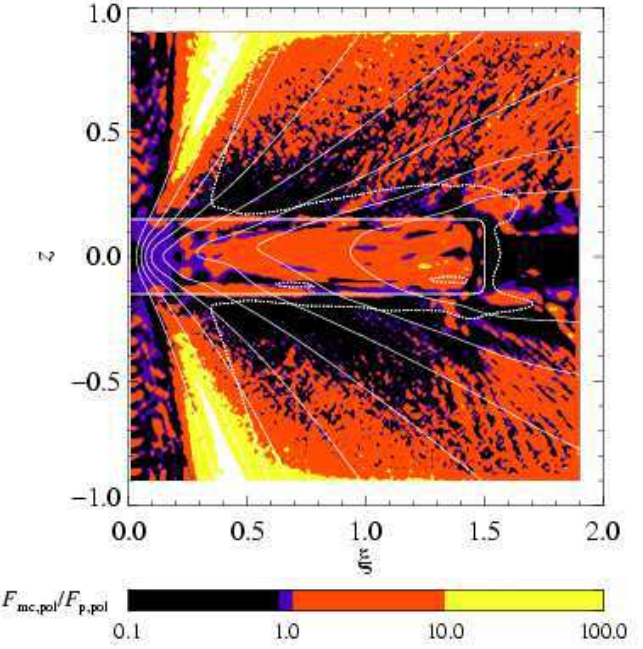}}
\caption[]{
The ratio of the poloidal magneto-centrifugal and pressure forces,
$|\FF_{\rm pol}^{\rm(mc)}|/|\FF_{\rm pol}^{\rm(p)}|$ as defined in Eq.~(\ref{PolForces}),
is shown with shades of grey, with larger values corresponding to lighter shades.
The maximum value is $|\FF_{\rm pol}^{\rm(mc)}|/|\FF_{\rm pol}^{\rm(p)}|\approx 411$.
Superimposed are the poloidal magnetic field lines.
The dashed line shows the
fast magnetosonic surface with respect to the poloidal field (cf.\ Fig.~\ref{Flito1}).
The parameters are as in the model of Fig.~\protect\ref{FRun7}.
Averaged over times $t=130...140$.
}\label{FRun7new_0.1_pforces}\end{figure}

\begin{figure}[t!]
\centerline{%
  \includegraphics[width=8.5cm]{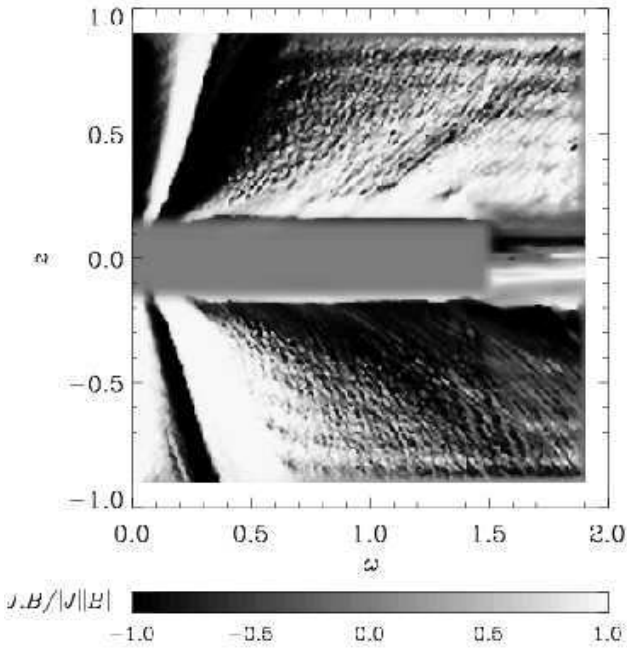}}
\caption[]{
The ratio $\JJ \cdot \BB / |\JJ|\, |\BB|$ in the corona
for the model with $\tau_{\rm star} = 0.01$ and $\beta=0.1$, shown in Fig.~\ref{FRun7}.
For a force-free magnetic field, $\JJ = C \BB$, this ratio is $\pm 1$.
Averaged over times $t=130...140$.
}\label{FRun7new01pffree}\end{figure}

\begin{figure}[t!]
\centerline{%
  \includegraphics[width=8.5cm]{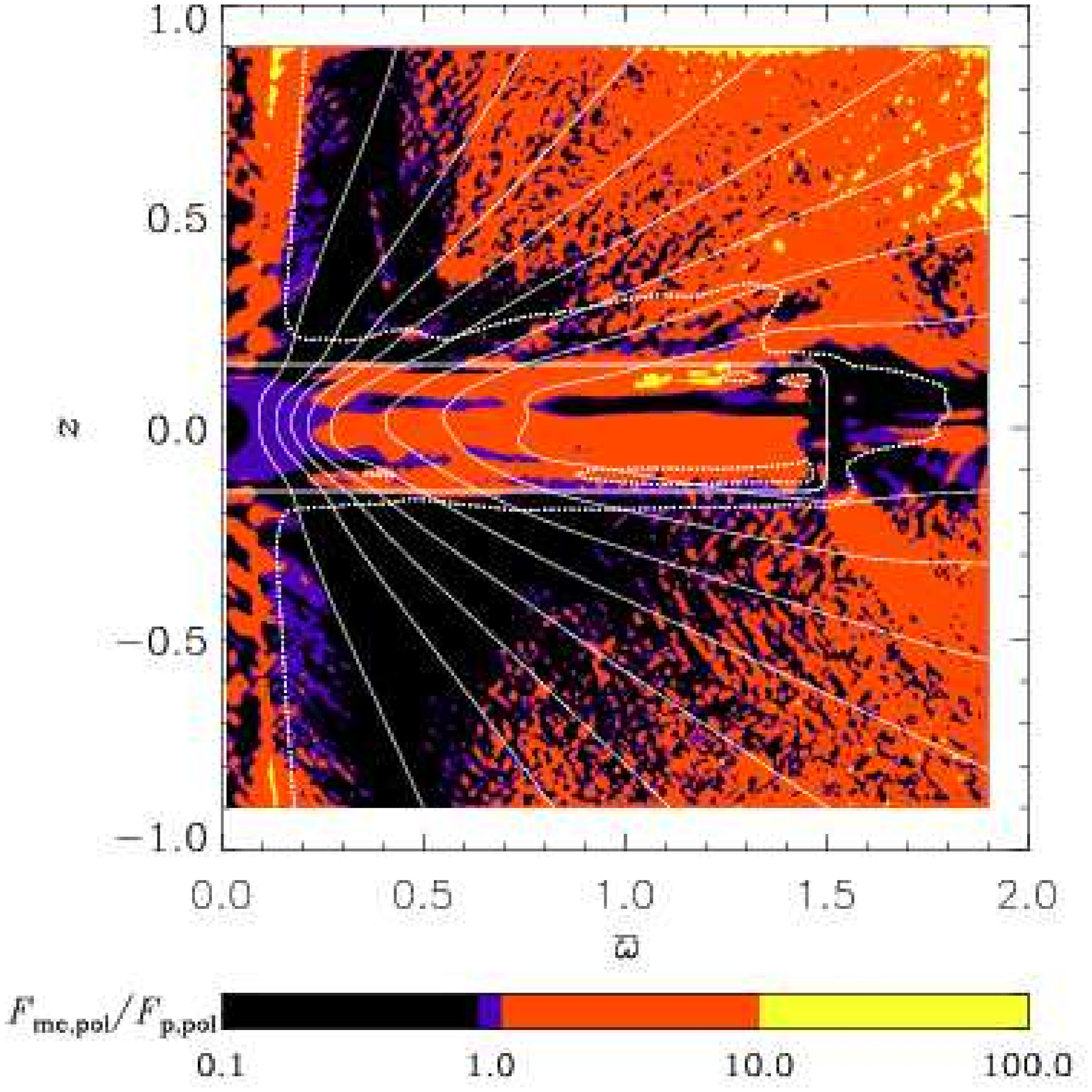}}
\caption[]{
As in Fig.~\protect\ref{FRun7new_0.1_pforces}, but for the model of
Fig.~\protect\ref{FRun7_0.02}, i.e.\ with larger entropy contrast, $\beta=0.02$.
Averaged over times $t=140...240$.
}\label{FRun7new_0.02_pforces}\end{figure}

The magnetic field geometry (e.g.,\ Fig.~\ref{FRun7}) is
such that for $\varpi > 0.1$,
the angle between the disc surface and the
field lines is less than $60^\circ$, reaching $\approx 20^\circ$ at
$\varpi \approx 1$--$1.5$, which is favourable for
magneto-centrifugal acceleration (Blandford \& Payne 1982;
Campbell 1999, 2000, 2001).
However, the Alfv\'en surface is so close to the disc surface in the outer parts
of the disc that acceleration there is mainly due to pressure gradient.
The situation is, however, different in the conical shell containing the fast wind.
As can be seen from Fig.~\ref{FRun7new01pangmom}, the Alfv\'en surface is
far away from the disc in that region and,
on a given field line,
the Alfv\'en radius is at least a few times larger than
the radius of the footpoint in the disc.
This is also seen in simulations of the magneto-centrifugally driven jets of
Krasnopolsky et al.\ (1999);
see their Fig.~4.
The lever arm of about $3$ is sufficient for magneto-centrifugal driving
to dominate.
As can be seen also from Fig.~\ref{Fpvaccall},
the flow at the polar angle $\vartheta=60^\circ$ is mainly accelerated by pressure gradient
near the disc surface
(where the Alfv\'en surface is close to the disc surface).
However, acceleration remains efficient out to at least $r=1$
within the conical shell at $\vartheta\approx30^\circ$. This can be seen in the upper and middle
panels of Fig.~\ref{Fpvaccall} (note that the conical shell is thinner and at a smaller $\vartheta$
in the model with larger
entropy contrast, and so it cannot be seen in this figure, cf.\ Fig.~\ref{FRun7_0.02}).
These facts strongly indicate that magneto-centrifugal acceleration dominates
within the conical shell.

Another indicator of magneto-centrifugal acceleration in the conical shell is the distribution
of angular momentum (see Fig.~\ref{FRun7new01pangmom}), which is significantly larger in the outer parts of
the conical shell than in the disc, which suggests that the magnetic field plays
an important r\^ole in the flow dynamics. We
show in Fig.~\ref{FRun7new_0.1_pforces} the ratio of the `magneto-centrifugal force'
to pressure gradient,
$|\FF_{\rm pol}^{\rm(mc)}|/|\FF_{\rm pol}^{\rm(p)}|$, where
the subscript `pol' denotes the poloidal components.
Here, the `magneto-centrifugal force' includes all terms in the
  poloidal equation of motion, except for the pressure gradient (but we ignore
  the viscous term and the mass production term, the latter being
  restricted to the disc only),
\EQ
\FF^{\rm(mc)}=\varrho\left(\Omega^2\vec{\varpi}-\vec{\nabla}\Phi\right)+\JJ\times\BB,
\label{PolForces}
\EN
and $\FF^{(p)}=-\vec{\nabla}p$.
The large values of the ratio in the conical shell confirm that
magneto-centrifugal driving is dominant there.
On the other hand, the pressure gradient is strong enough in the outer parts
of the disc to shift the Alfv\'en surface close to the disc surface, leading to
pressure driving. This is also discussed in Casse \& Ferreira (2000b) who
point out that, although the criterion of Blandford \& Payne (1982) is fulfilled,
thermal effects can be strong enough to lead to pressure driving.
  According to Ferreira (1997), a decrease of the total poloidal
  current $I_{\rm pol} = 2 \pi \varpi B_\varphi / \mu_0$ along a field line
  is another indicator of magneto-centrifugal acceleration.
  We have compared the poloidal current $I_{\rm pol}^{\rm(top)}$ at the
  Alfv\'en point, or at the top of our box if the Alfv\'en point is outside the
  box, with the poloidal current $I_{\rm pol}^{\rm(surf)}$ at the disc surface,
  and find that outside the conical shell this ratio is typically $\ge 0.8$,
  while along
  the field line that leaves the box at $(\varpi,z)=(0.6,1)$, we get
  $I_{\rm pol}^{\rm(top)} / I_{\rm pol}^{\rm(surf)} \approx 0.18$, i.e.\ a
  strong reduction, which confirms that magneto-centrifugal acceleration
  occurs inside the conical shell.
  We note, however, that the changing sign of $B_\varphi$ and therefore of
  $I_{\rm pol}$ makes this
  analysis inapplicable in places, and the distribution of angular
  momentum (Fig.~\ref{FRun7new01pangmom}) gives a much clearer picture.

As further evidence of a significant contribution from
magneto-centrifugal driving in the conical shell, we show in Fig.~\ref{FRun7new01pffree}
that the magnetic field is close to a force-free configuration in regions where
angular momentum is enhanced, i.e.\ in the conical shell and in the corona surrounding
the outer parts of the disc. These are the regions where the Lorentz force contributes
significantly to the flow dynamics, so that the magnetic field performs work and therefore relaxes
to a force-free configuration.
The radial variation in the sign of the current helicity $\JJ\cdot\BB$ is due to
a variation in the sign of the azimuthal magnetic field and of the current
density.
Such changes originate from the disc where they
imprint a corresponding variation in the sign of the
angular momentum constant, see Eq.~(\ref{ell}).
These variations are then carried along magnetic lines into the corona.
The locations where the azimuthal field reverses are still relatively
close to the axis, and there the azimuthal field relative to the poloidal field
is weak compared to
regions further away from the axis.

Pressure driving is more important if the entropy contrast between
the disc and the corona is larger (i.e.\ $\beta$ is smaller):  the white conical shell
in Fig.~\ref{FRun7new_0.1_pforces}
indicative of stronger magneto-centrifugal driving shifts to larger heights for $\beta=0.02$,
as shown in Fig.~\ref{FRun7new_0.02_pforces}. We note, however, that there
are periods when magneto-centrifugal driving is dominant even in this
model with higher entropy contrast, but pressure driving dominates in the
time averaged picture (at least within our computational domain).

\begin{figure}[t!]
\centerline{%
  \includegraphics[width=8.9cm]{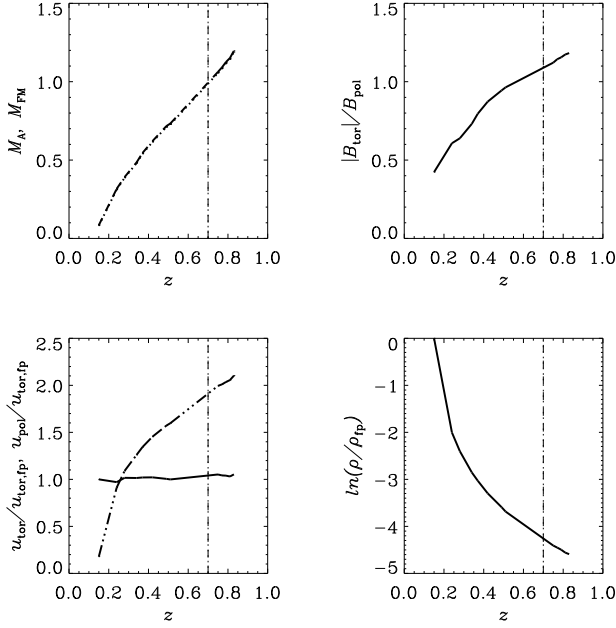}}
\caption[]{
Flow parameters as functions of height $z$, along the field line with its footpoint at
$(\varpi_{\rm fp},z_{\rm fp})=(0.17,0.15)$,
for the model of Fig.~\ref{FRun7}.
Upper left:
poloidal
Alfv\'en Mach number, $M_{\rm A}\equiv u_{\rm pol}/v_{\rm A,pol}$
(dashed), and
a similar quantity that includes the poloidal Alfv\'en speed
as well as the sound speed (fast magnetosonic Mach number with respect to
the poloidal field),
$M_{\rm FM}\equiv u_{\rm pol}/(c_{\rm s}^2+v_{\rm A,pol}^2)^{1/2}$
(dotted)
-- the two curves are practically identical;
upper right: ratio of toroidal ($B_{\rm tor}$) and poloidal ($B_{\rm pol}$)
magnetic fields;
lower left: toroidal velocity $u_{\rm tor}$ (solid) and poloidal velocity
$u_{\rm pol}$ (dash-3dots), in units of the toroidal velocity at the footpoint,
$u_{\rm tor,fp}$;
lower right: density $\varrho$ in units of the density at the footpoint,
$\varrho_{\rm fp}$.
The position of
the Alfv\'en (and fast magnetosonic) point
on the field line is
indicated by the vertical line.
This figure is useful to compare with Fig.\ 3 of Ouyed et al.\ (1997).
}\label{Fpklall}\end{figure}

We show in Fig.~\ref{Fpklall} the variation of several quantities along
a magnetic field line that has its footpoint at the disc
surface at $(\varpi,z)=(0.17,0.15)$ and lies around the conical shell.
Since this is where magneto-centrifugal driving is still dominant, it is useful
to compare Fig.~\ref{Fpklall} with Fig.\ 3 of Ouyed et al.\ (1997), where a
well-collimated magneto-centrifugal jet is studied.
Since our outflow is collimated only weakly within our computational domain,
the quantities are plotted against height $z$, rather than $z/\varpi$ as in
Ouyed et al.\ (1997) ($z/\varpi$ is nearly constant along a field line for weakly collimated
flows, whereas approximately $z/\varpi \propto z$ along a magnetic line for well-collimated flows).
The results are qualitatively similar, with the main difference that
the fast magnetosonic surface in our model almost coincides with the Alfv\'en surface
in the region around the conical shell where the outflow is highly supersonic.
Since we include finite diffusivity, the curves in Fig.~\ref{Fpklall} are smoother
than in Ouyed et al.\ (1997), who consider ideal MHD.
A peculiar feature of the conical shell is that the flow at $z \la 1$ is sub-Alfv\'enic
but strongly supersonic.
The fast magnetosonic surface is where the poloidal
  velocity $v_{\rm pol}$ equals the fast magnetosonic speed for the
  direction parallel to the field lines,
  \begin{equation}
    v_{\rm pol}^2 ={1\over2}\left(
    {c_{\rm s}^2{+}v_{\rm A}^2
          + \sqrt{(c_{\rm s}^2{+}v_{\rm A}^2)^2
                  - 4c_{\rm s}^2v_{\rm A,pol}^2}}\right),
\end{equation}
with $v_{\rm A,pol}$ the
Alfv\'en speed from the poloidal magnetic field. This surface has the same
overall shape as the fast magnetosonic surface with respect to the poloidal field,
albeit in some cases it has a
somewhat larger opening angle around the conical shell and is located further
away from the disc in regions where the toroidal Alfv\'en speed is
enhanced.

\subsection{Lagrangian invariants} \label{Lagr}

Axisymmetric ideal magnetized outflows are governed by five
Lagrangian invariants,
the flux ratio,
\EQ
k = \varrho u_z/B_z = \varrho u_\varpi/B_\varpi,
\EN
the angular velocity of magnetic field lines,
\EQ
\widetilde\Omega = \varpi^{-1} (u_\varphi - k B_\varphi/\varrho),
\EN
the angular momentum constant,
\EQ
\ell=\varpi u_\varphi-\varpi B_\varphi/(\mu_0 k), \label{ell}
\EN
the Bernoulli constant,
\EQ
e=\half\uu^2+h+\Phi-\varpi\widetilde\Omega B_\varphi/(\mu_0 k).
\EN
and specific entropy $s$ (which is a prescribed function of position in our
model).
In the steady state, these five quantities
are conserved along field lines, but vary from one magnetic field line to
another (e.g.,\ Pelletier \& Pudritz 1992; Mestel 1999),
i.e.\ depend on the magnetic flux
within a magnetic flux surface, $2\pi a$, where $a=\varpi A_\varphi$ is
the flux function whose contours represent poloidal field lines.

\begin{figure}[t!]
\centerline{%
  \includegraphics[width=8.5cm]{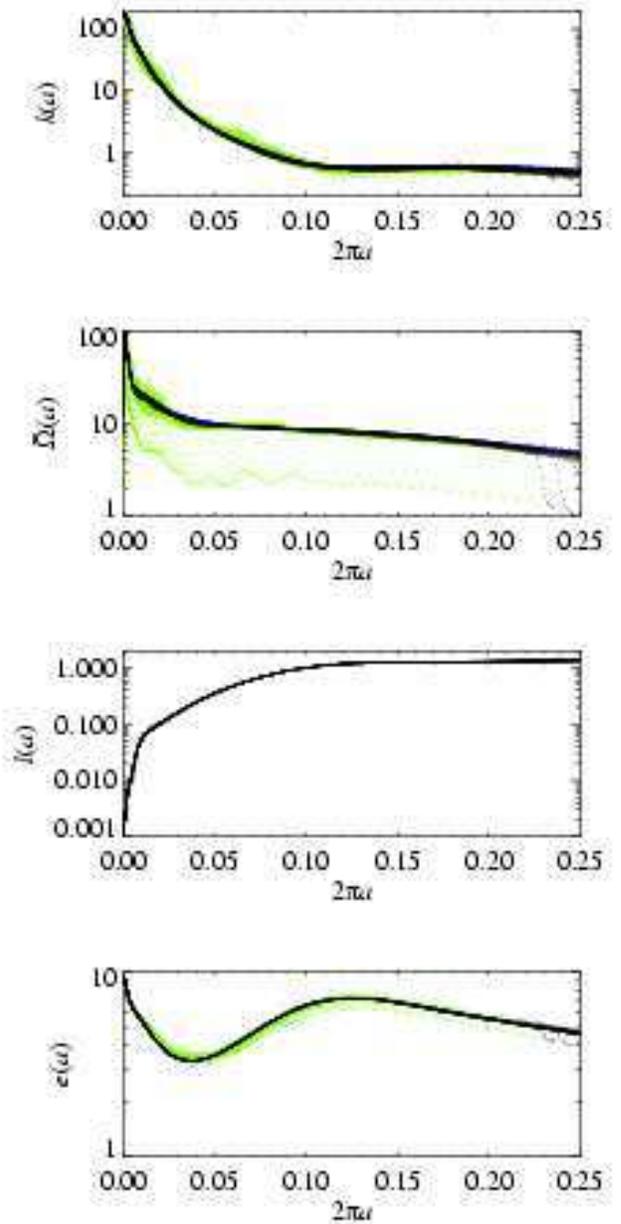}}
\caption[]{
The four Lagrangian invariants $k(a)$, $\widetilde\Omega(a)$, $\ell(a)$, and $e(a)$
as a function of the flux function $a$
for the model of Fig.~\ref{FRRRRun3c}.
All points in the domain $0<\varpi<8$, $0.2\leq z\leq30$
are shown (provided $0<a\le0.25$).
The dots deviating from well-defined curves mostly originate in $8\la z\leq30$.
Note that specific entropy as fifth Lagrangian invariant is trivially
conserved in our model, because entropy is spatially constant
throughout the corona and temporally fixed.
}\label{Fpkl}\end{figure}

We show in Fig.~\ref{Fpkl} scatter plots of $k(a)$, $\widetilde\Omega(a)$, $\ell(a)$, and $e(a)$
for the model of Fig.~\ref{FRRRRun3c}.
Points from the region $0.2\le z\la 8$ collapse into a single line, confirming that
the flow in the corona is nearly ideal.%
\footnote{
  If we restrict ourselves to $0.2\le z\la 8$, about $90\%$ of the points
  for $\widetilde{\Omega}(a)$ are
  within $\pm 10\%$ of
  the line representing $z=4$;
  for the other three invariants, this percentage is at least $97\%$.
  In a larger domain,
  $0.2\le z\la 30$,
  these percentages drop to
  about $80\%$ for $k(a)$ and $60\%$ for $\widetilde\Omega(a)$;
  for $\ell(a)$ and $e(a)$, however, they remain greater than $90\%$.
}
This is not surprising since
the magnetic Reynolds number is much greater than
unity in the corona
for the parameters adopted here.
For $8\la z\le 30$, there are departures from perfect MHD; in particular, the
angular velocity of magnetic field lines, $\widetilde\Omega$,
is somewhat decreased in the upper parts of the domain
(indicated by the vertical scatter in the data points).
This is plausibly due to the finite magnetic diffusivity which still
allows matter to slightly lag behind the magnetic field.
As this lag accumulates along a stream line, the departures increase with
height $z$.
  Since this is a `secular' effect only, and accumulates with height, we locally
  still have little variation of $k$ and $\widetilde\Omega$, which explains why
  magneto-centrifugal acceleration can operate quite efficiently.

The corona in our model has (turbulent) magnetic diffusivity comparable to that in the disc.
This is consistent with, e.g.,\ Ouyed \& Pudritz (1999) who argue that turbulence
should be significant in coronae of accretion discs.
Nevertheless, it turns out that ideal MHD is a reasonable approximation for the corona (see
Fig.~\ref{Fpkl}), but not for the disc. Therefore, magnetic diffusivity is
physically significant in the disc and insignificant in the corona (due to
different velocity and length scales involved), as in most models of disc
outflows (see Ferreira \& Pelletier 1995 for a discussion). Thus, our model
confirms this widely adopted idealization.

\section{Toward more realistic models}
\label{FinalModel}

The models presented so far have some clear deficiencies when compared
with characteristic features relevant to protostellar discs. The first deficiency concerns
the relative amount of matter accreted onto the central object compared
with what goes into the wind. A typical figure from the models presented
above was that as much as 30\% of all the matter released in the disc
goes into the wind and only about 70\% is accreted by the central
object. Earlier estimates (e.g.,\ Pelletier \& Pudritz 1992) indicate that only about
10\% of the matter joins the wind. Another possible deficiency
is the fact that in the models presented so far the low-temperature region extends
all the way to the stellar surface whilst in reality the cool disc breaks
up close to the star because of the stellar magnetic field. Finally,
the overall temperature of the disc is generally too high compared with
real protostellar discs which are known to have typical temperatures of
about a few thousand Kelvin.

The aim of this section is to assess the significance of various
improvements in the model related to the above mentioned characteristics.
We consider the effect of each of them separately by improving the model step by step.

\subsection{A larger sink} \label{LS}

As discussed in Sects.~\ref{NMO} and \ref{MO}, properties of the outflow,
especially of the nonmagnetic ones, are sensitive to the parameters of the
central sink. It is clear that a very efficient sink would completely inhibit
the outflow near the axis. On the other hand, a magnetosphere of the central
object can affect the sink efficiency by channelling the flow along the stellar
magnetic field (Shu et al.\ 1994; see also Fendt \& Elstner 1999, 2000).
Simulating a magnetosphere turned out
to be a difficult task and some preliminary attempts proved unsatisfactory.

\begin{figure}[t!]
\centerline{
  \includegraphics[width=8.5cm]{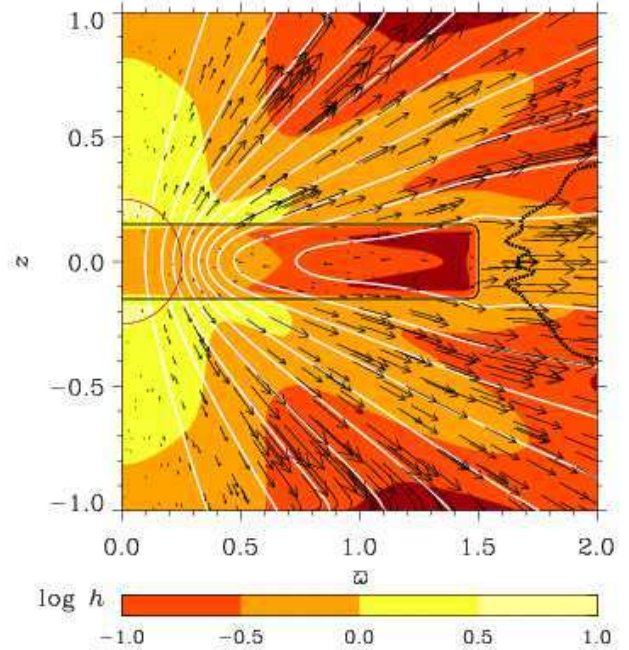}}
\caption[]{
As in Fig.~\protect\ref{FRun7}, but with a mass sink at the centre which is
five times larger, $r_0 = 0.25$, and
$\tau_{\rm star}=0.112$.
Note the absence of outflows along the axis
and a larger opening angle of the conical shell,
which crosses $z=\pm1$ at $\varpi\approx1.2$.
Averaged over times $t=500 \dots 530$, $\beta=0.1$.
}\label{FRun7new_sink2_tau2_pnew}\end{figure}

Instead, we have considered a model with a geometrically larger sink,
and illustrate it in Fig.~\ref{FRun7new_sink2_tau2_pnew}. Here, $r_0=0.25$ in
Eqs~(\ref{xistar}) and (\ref{gravpot}), which is five times larger than the sink used in
our reference model of Fig.~\ref{FRun7}. The relaxation time of the sink
in Eq.~(\ref{qminus}), $\tau_{\rm star}$, was rescaled in proportion to
the free-fall time at $r_0$ as $\tau_{\rm star}\simeq r_0/v_{\rm ff}\propto r_0^{3/2}$,
which yields $\tau_{\rm star}=0.112$. The resulting mass loss rate into the
wind is $\dot M_{\rm wind}\approx0.8$, which corresponds to
$1.6\times10^{-7}\,M_\odot\yr^{-1}$ in dimensional units. Although this is indeed smaller
than the value for the reference model ($6\times10^{-7}\,M_\odot\yr^{-1}$),
the overall mass released from the disc is also smaller, resulting in
a larger fraction of about 40\% of matter that goes into the wind;
only about 60\% is accreted by the central object.
As we show below,
however, larger accretion fractions can more easily be achieved by making
the disc cooler.
We conclude that even a sink as large as almost twice the disc half-thickness does
not destroy the outflow outside the inner cone. However, the outflow along the axis
is nearly completely suppressed.

\subsection{Introducing a gap between the disc and the sink}   \label{IGBDS}
\begin{figure}[t!]
\centerline{
  \includegraphics[width=8.5cm]{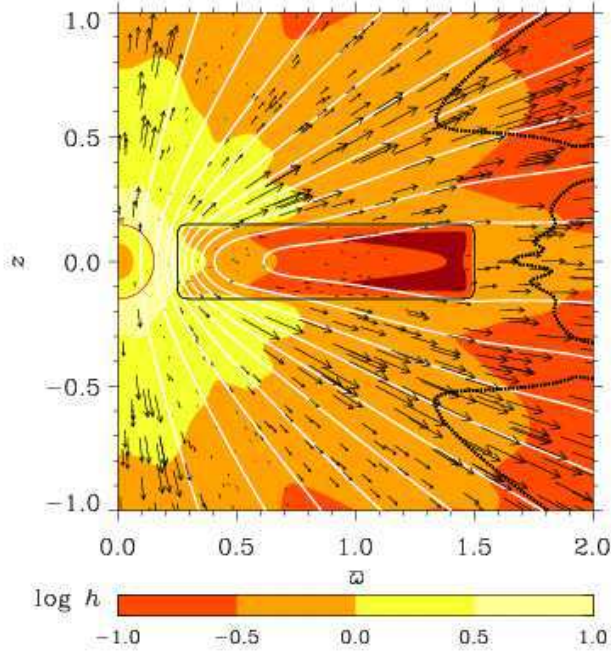}}
\caption[]{
As in Fig.~\protect\ref{FRun7}, but with a gap between the sink and the disc
with the disc inner radius at $\varpi=0.25$.
The radius of the stellar mass sink is three times larger, $r_0 = 0.15$,
and $\tau_{\rm star}=0.052$.
Outflow is absent along field lines passing through the gap.
Averaged over times $t=400 \dots 660$, $\beta=0.1$ in both the disc and the sink.
}\label{FRun7new_large_pnew}\end{figure}

In real accretion discs around protostars the disc terminates at some
distance away from the star. It would therefore be unrealistic to let
the disc extend all the way to the centre. Dynamo action in all our models
is restricted to $\varpi\geq0.2$, and
now we introduce an inner disc boundary for the region of
lowered entropy as well. In Fig.~\ref{FRun7new_large_pnew} we present such a
model where the $\xi_{\rm disc}(\vec{r})$ profile is terminated at $\varpi=0.25$.
This inner disc radius affects then not only the region of lowered entropy, but also
the distributions of $\alpha$, $\eta_{\rm t}$, $\nu_{\rm t}$ and $q_{\varrho}^{\rm disc}$.
The radius of the mass sink was chosen to be $r_0=0.15$,
i.e.\ equal to the disc semi-thickness. The correspondingly adjusted value
of $\tau_{\rm star}$ is 0.052.
The entropy in the sink
is kept as low as that in the disc.

As can be seen from Fig.~\ref{FRun7new_large_pnew}, the gap in the disc
translates directly into a corresponding gap in the resulting outflow
pattern. At the same time, however, only about 20\% of the released disc material
is accreted by the sink and the rest is ejected into the wind.
A small fraction of the mass that accretes toward the sink reaches
the region near the axis at some distance away from the origin and can
then still be
ejected as in the models without the gap.
Note that our figures (e.g.,  Fig.~\ref{FRun7new_large_pnew})
show the velocity rather than the azimuthally integrated mass flux density
(cf.\ Fig.~\ref{stream}); the relative magnitude of the latter is much smaller,
and so the mass flux from the sink region is not significant.

\subsection{A cooler disc}

We now turn to the discussion of the case with a cooler disc. Numerical
constraints prevent us from making the entropy gradient between disc and
corona too steep. Nevertheless, we were able to reduce the value of $\beta$
down to 0.005, which is 20 times smaller than the value used for the
reference model in Fig.~\ref{FRun7}. The $\beta$ value for the sink was
reduced to 0.02; smaller values proved numerically difficult.

\begin{figure}[t!]
\centerline{
  \includegraphics[width=8.5cm]{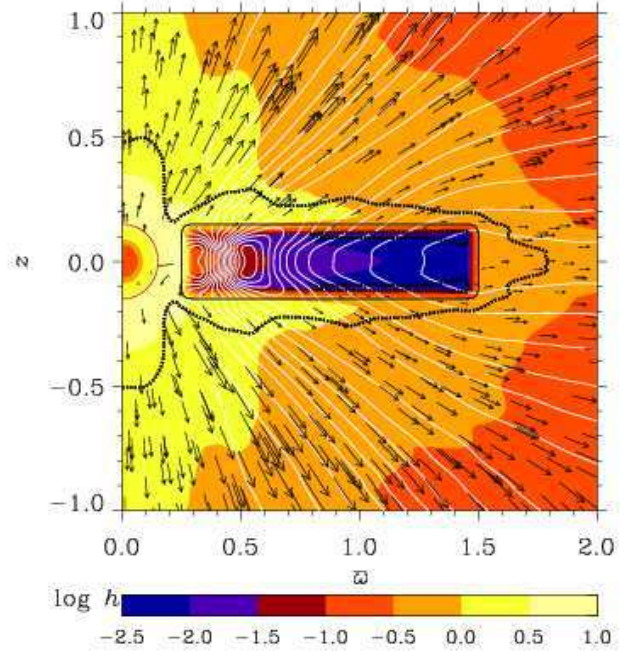}}
\caption[]{
As in Fig.~\protect\ref{FRun7new_large_pnew},
but with $\beta = 0.005$ in the disc and
$\beta=0.02$ in the sink.
Averaged over times $t=150 \dots 280$
when the overall magnetic activity in the
disc is relatively low.
}\label{FRun7new_cold_lowosc_pnew}\end{figure}

\begin{figure}[t!]
\centerline{
  \includegraphics[width=8.5cm]{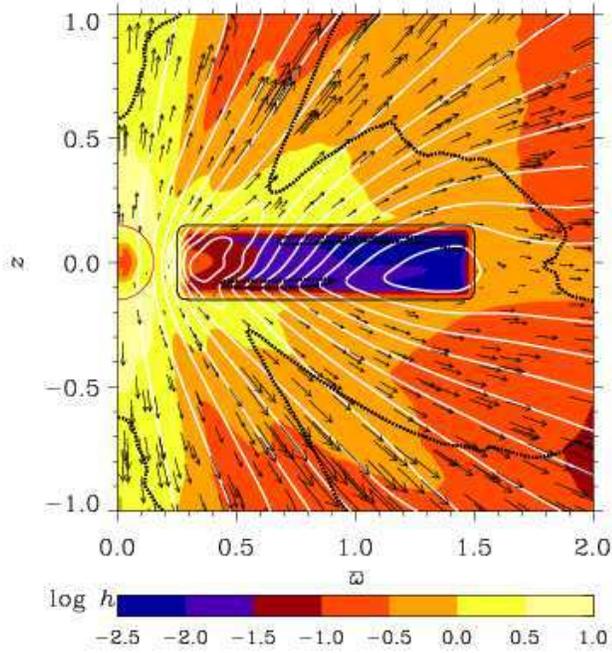}}
\caption[]{
As in Fig.~\protect\ref{FRun7new_cold_lowosc_pnew},
but averaged over a period of enhanced magnetic activity
in the disc,
$t=300 \dots 320$.
A conical shell develops which crosses $z=\pm1$ at $\varpi\approx0.6$.
}\label{FRun7new_cold_highosc_pnew}\end{figure}

The value of $\beta = 0.005$ results in a disc temperature of
$3\times10^3$\,K in the outer parts and $1.5\times10^4\,$K in
the inner parts. As in Sect.~\ref{IGBDS}, the disc terminates
at an inner radius of $\varpi=0.25$. At $\varpi\approx0.5$, the density
in dimensional units is about $10^{-9}\,$g\,cm$^{-3}$, which is
also the order of magnitude found for protostellar discs.
The resulting magnetic field and outflow geometries are shown in
Figs.~\ref{FRun7new_cold_lowosc_pnew} and \ref{FRun7new_cold_highosc_pnew}.

A characteristic feature of models with a cooler disc is a more
vigorous temporal behaviour with prolonged episodes of reduced overall
magnetic activity in the disc during which the Alfv\'enic surface
is closer to the disc surface, and phases of enhanced magnetic activity
where the Alfv\'enic surface has moved further away.
Figures~\ref{FRun7new_cold_lowosc_pnew} and \ref{FRun7new_cold_highosc_pnew}
are representative of these two states.
It is notable that the structured outflow of the type seen in the reference
model
occurs in states with strong magnetic field and disappears during periods with
weak magnetic field.

Another interesting property of the cooler discs is that now a smaller
fraction of matter goes into the wind (10--20\%), and 80--90\% is
accreted by the central object, in a better agreement with the estimates
of Pelletier \& Pudritz (1992).

We note in passing that channel flow solutions typical of two-dimensional
simulations of the magneto-rotational instability (Hawley \& Balbus 1991)
are generally absent in the simulations presented here. This is because
the magnetic field saturates at a level close to equipartition between
magnetic and thermal energies. The vertical wavelength of the instability
can then exceed the half-thickness of the disc. In some of our
simulations, indications of channel flow behaviour still can be seen. An example
is Fig.~\ref{FRun7new_cold_lowosc_pnew} where the magnetic energy is weak
enough so that the magneto-rotational instability is not yet suppressed.

According to the Shakura--Sunyaev prescription, turbulent viscosity and
magnetic diffusivity are reduced in a cooler disc because of the smaller
sound speed (cf. Eq.~(\ref{nut})).
Since in the corona the dominant contributions to the artificial advection
viscosity $\nu_{\rm adv}$ come from the poloidal velocity and poloidal Alfv\'en
speed (and not from the sound speed), $c_\nu^{\rm adv}$ has to be reduced. Here
we choose $\alpha_{\rm SS}=0.004$ and reduce $c_\nu^{\rm adv}$ by a factor of
$10$ to $c_\nu^{\rm adv}=0.002$.
Since we do not explicitly parameterize the turbulent magnetic diffusivity
$\eta_{\rm t}$ with the sound speed (cf.\ Eq.~(\ref{eeta})), also
$\eta_{{\rm t}0}$ needs to be decreased, together with the background
diffusivity $\eta_0$. We choose here values that are $25$ times smaller compared
to the previous runs, i.e.\ $\eta_{{\rm t}0}=4\times10^{-5}$ and
$\eta_0=2\times10^{-5}$, which corresponds to $\alpha_{\rm SS}^{(\eta)}$
ranging between $0.001$ and $0.007$.
With this choice of coefficients, the total viscosity and magnetic diffusivity
in the disc have comparable values of a few times $10^{-5}$.

\begin{figure}[t!]
\centerline{
  \includegraphics[width=8.5cm]{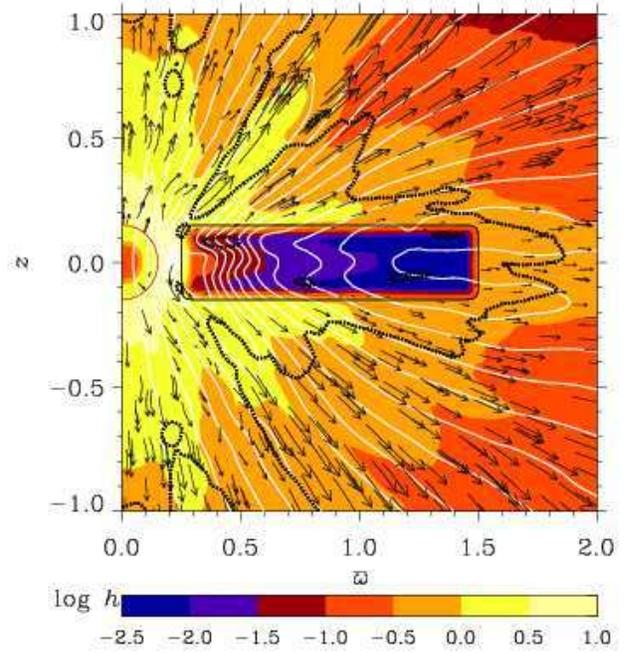}}
\caption[]{
As in Figs.~\protect\ref{FRun7new_cold_lowosc_pnew} and \ref{FRun7new_cold_highosc_pnew},
but with $\eta_0=2\times10^{-5}$,
$\eta_{{\rm t}0}=4\times10^{-5}$,
$c_\nu^{\rm adv}=0.002$,
and $\alpha_0=-0.15$.
Averaged over times $t=230 \dots 236$ (where the magnetic energy is enhanced),
$\alpha_{\rm SS}=0.004$.
}\label{FRun7new_cold_st_pnew}\end{figure}

The effect of reduced viscosity and magnetic diffusivity is shown in
Fig.~\ref{FRun7new_cold_st_pnew}. Features characteristic of the
channel flow solution are now present,
because the vertical wavelength of the magneto-rotational instability is here
less than the half-thickness of the disc.

\section{Discussion}   \label{Discu}

If the disc dynamo is sufficiently strong,
our model develops a clearly structured outflow
which is
fast, cool and rarefied within a conical shell near the rotation axis
where most of the angular momentum and magnetic energy is carried,
and is slower, hotter and denser in the region around the axis
as well as in the outer parts of the domain.
The slower outflow is driven mostly by the entropy
contrast between the disc and the corona, but the faster wind
within the conical shell is mostly driven magneto-centrifugally.
Without a central mass sink, the flow near the axis
is faster, but otherwise the flow structure is similar to that with the sink.

The half-opening angle of the cone with hot, dense gas around the axis is
about $20^\circ$--$30^\circ$; this quantity somewhat changes with model parameters but remains
close to that range.
The outflow in our models does not show any signs of collimation. It should be noted,
however, that not all outflows from protostellar discs are
actually collimated, especially not at such small distance
from the source. An example is the
Becklin--Neugebauer/Kleinmann--Low (BN/KL) region in the Orion Nebula
(Greenhill et al.\ 1998), which has a conical outflow with a half-opening
angle of $30^\circ$ out to a distance of 25--60\,AU from its origin.
Therefore, collimation within a few AU (the size of our computational domain)
is expected to be only weak.

The region around the fast, cool and rarefied conical shell seen in Fig.~\ref{FRun7}
is similar to the flow structure reported by Krasnopolsky et al.\ (1999);
see their Fig.~1. In their model, however, the thin axial jet was caused by
an explicit injection of matter from the inner parts of the disc which was
treated as a boundary. In our reference model the fast outflow is
sub-Alfv\'enic because of the presence of
a relatively strong poloidal field, whereas in Krasnopolsky et al.\ (1999)
the outflow becomes super-Alfv\'enic at smaller heights.
Outside the conical shell the outflow is mainly pressure driven,
even though the criterion of Blandford \& Payne (1982) is fulfilled.
However, as Casse \& Ferreira (2000b) pointed out,
pressure driven outflows might
dominate over centrifugally driven outflows if thermal
effects are strong enough.

In our model, matter
is replenished in the resolved disc in a self-regulatory manner where and when
needed.
We believe that this is an improvement in comparison to the models of
Ouyed \& Pudritz (1997a, 1997b, 1999) and Ustyugova et al.\ (1995),
where mass inflow is prescribed as a boundary condition at the base of the
corona.
If we put $q_\varrho^{\rm disc}=0$ in
Eqs~(\ref{Cont}) and (\ref{momentum}), the disc mass soon drops to low values and
the outflow ceases. This is qualitatively the same behaviour as in the models of,
e.g.,\ Kudoh et al.\ (1998).

We should stress the importance of finite magnetic diffusivity
in the disc: although poloidal velocity and poloidal magnetic field are well aligned in
most of the corona, dynamo action in the disc is only possible in the presence of
finite magnetic diffusivity, and the flow can enter the corona only by crossing
magnetic field lines in the disc.

An outflow occurs in the presence of both dipolar and quadrupolar type magnetic fields,
even though fields with dipolar symmetry seem to be more efficient
in magneto-centrifugal driving (cf.\ von Rekowski et al.\ 2000).
The effects of the magnetic parity on the outflow structure deserves further analysis.

The dynamo active accretion disc drives a significant outward Poynting flux in our model.
Assuming that this applies equally to protostellar and AGN discs, this
result could be important for understanding the origin of seed magnetic
fields in galaxies and galaxy clusters; see Jafelice \& Opher (1992) and
Brandenburg (2000) for a discussion. We note, however, that the pressure
of the intracluster gas may prevent the magnetized plasma from active
galactic nuclei to spread
over a significant volume (Goldshmidt \& Rephaeli 1994).

Our model can be improved in several respects.
In many systems, both dynamo-generated and external magnetic fields may be
present, so a more comprehensive model should include both.
We used an $\alpha^2\Omega$ dynamo to parameterize magnetic field
generation in the disc because we have restricted ourselves to
axisymmetric models.
As argued by Brandenburg (1998), dynamo action of turbulence which is
driven by the
magneto-rotational instability can be roughly described as an $\alpha^2\Omega$ dynamo.
But this parameterization can be relaxed in
three-dimensional simulations
where one may expect that turbulence will be generated to drive dynamo action.
Such simulations will be discussed elsewhere (von Rekowski et al.\ 2002).

Since our model includes angular momentum transport by both viscous and
magnetic stresses, it is natural that the accreted matter is eventually
diverted into an outflow near the axis; this is further facilitated by our
prescribed entropy gradient at the disc surface. We believe
that this picture is physically well motivated (Bell \& Lucek 1995), with the
only reservation that we do not incorporate the (more complicated) physics of
coronal heating and disc cooling, but rather parameterize it with a fixed entropy contrast.
We include a mass sink at the centre which
could have prevented the outflow, and indeed the sink strongly affects nonmagnetized
outflows. We have shown, however,
that the magnetic field can efficiently shield the sink and thereby support a vigorous
disc wind.

The assumption of a prescribed entropy
distribution is a useful tool to control the size of the disc and to parameterize
the heating of the disc corona.  However, it
should be relaxed as soon as the disc physics can be described more fully.
The energy equation, possibly with radiation transfer, should be
included. This would lead to a self-consistent entropy
distribution and would admit the deposition of viscous
and Ohmic heat in the outflow.
In the simulations by von Rekowski et al.\ (2002), entropy is evolved.

We believe that a mass source is
a necessary feature of any model of this kind if one wishes to obtain a steady
state.
In the present paper the mass source is distributed throughout the whole
disc to represent replenishment of matter from the midplane
of the disc.
Alternatively, a mass
source could be located near to or on the domain boundary.

\begin{acknowledgements}
  We are grateful to C.G.~Campbell, R.~Henriksen, J.~Heyvaerts, R.~Ouyed and R.E.~Pudritz
  for fruitful discussions.
We acknowledge many useful comments of the anonymous referee.
 This work was partially supported by PPARC
  (Grants PPA/G/S/1997/00284 and 2000/00528) and the Leverhulme Trust (Grant
  F/125/AL).
Use of the PPARC supported supercomputers in St Andrews and Leicester
is acknowledged.
\end{acknowledgements}


\label{lastpage}

\vfill\bigskip\noindent{\it
$ $Id: wind.tex,v 1.105.2.28 2002/11/20 17:45:57 brandenb Exp $ $}

\end{document}